\documentclass[sn-basic]{sn-jnl}% Basic Springer Nature Reference Style/Chemistry Reference Style
\usepackage{amsmath}
\usepackage{graphicx}
\usepackage{natbib}
\usepackage{url} % not crucial - just used below for the URL 
\usepackage{amsfonts}
\usepackage{natbib,hyperref}
\usepackage{subcaption}
\usepackage{url} % not crucial - just used below for the URL 
\usepackage{cleveref}
\usepackage{xcolor}
\usepackage{float}
\usepackage{babel} 
\renewcommand{\figurename}{Figure}

\jyear{2022}%

\theoremstyle{thmstyleone}%
\theoremstyle{thmstyletwo}%

\theoremstyle{thmstylethree}%

\newtheorem{condition}{Condition}
\newtheorem{thm}{Theorem}

\DeclareMathOperator*{\argmin}{argmin}

\begin{document}

\title[Tree-based boosting with functional data]{Tree-based boosting with functional data}


\author*[]{\fnm{Xiaomeng} \sur{Ju}}\email{xiaomeng.ju@stat.ubc.ca}

\author[]{\fnm{Mat\'ias} \sur{Salibi\'an-Barrera}}\email{matias@stat.ubc.ca}

\affil[]{\orgdiv{Department} of Statistics, \city{Vancouver}, \state{BC}, \country{Canada}}

%\affil[2]{\orgdiv{Department}, \orgname{Organization}, \orgaddress{\street{Street}, \city{City}, \postcode{10587}, \state{State}, \country{Country}}}

%\affil[3]{\orgdiv{Department}, \orgname{Organization}, \orgaddress{\street{Street}, \city{City}, \postcode{610101}, \state{State}, \country{Country}}}


\abstract{In this article we propose a boosting algorithm for regression with functional explanatory variables and scalar responses. The algorithm uses decision trees constructed with multiple projections as the ``base-learners", which we call ``functional multi-index trees". We establish identifiability conditions for these trees and introduce two algorithms to compute them. 
%:  one  finds optimal projections over the entire tree, while the other one searches for a single optimal projection at each split.  

We use numerical experiments to investigate the performance of our method and compare it with several linear and nonlinear regression estimators, including recently proposed nonparametric and semiparametric functional additive estimators.  Simulation studies show that the proposed method is consistently among the top performers, whereas the performance of 
existing alternatives
can vary substantially across different settings. In a real example, we apply our method to predict electricity demand using price curves and show that our estimator provides better predictions compared to its competitors, especially when one adjusts for seasonality. }


\maketitle

\section{Introduction}
Functional data consist of a sample of functional variables,  which are often smooth functions measured on a discrete grid.  In this article, we study regression problems  with a functional covariate $X$ and a scalar response $Y$, also called ``scalar-on-function" regression models,  where the main goal is to estimate a function $F: X \rightarrow Y$ in order to make predictions for future observations of $X$.

Functional data are intrinsically infinite dimensional, and that poses  challenges that vary with how the regression model is defined. Assuming that $X \in L^2(\mathcal{I})$ are square-integrable functions defined over an interval $\mathcal{I}$,  the linear functional regression model is defined as $F(X) = \mu + \langle X, \beta \rangle,$ where $\beta \in L^2(\mathcal{I})$ is the functional regression coefficient and $ \langle X, \beta \rangle$ denotes the usual inner-product associated with the Hilbert space $L^2(\mathcal{I})$.  Since $\beta$ is infinite dimensional, its estimation requires some form of dimension reduction.  A very common way of fitting this model is to represent $\beta$ in some basis system (most often splines or functional principal components (FPCs)), and control its smoothness by regularizing the expansion, either through truncating the number of basis functions or by adding roughness penalties. A number of proposals in the literature explored different  basis expansions and 
regularization strategies, for example \cite{cardot2003spline}, 
\cite{hall2007methodology},
\cite{reiss2007functional}, and \cite{zhao2012wavelet}. 
For non-Gaussian responses, several authors studied the generalized functional linear model $g\left(F(X)\right) = \mu + \langle X, \beta \rangle$ for a known link function~$g$, see, e.g. \cite{james2002generalized}, \cite{cardot2005estimation}, \cite{muller2005generalized}, and \cite{dou2012estimation}.  

A well-specified linear model typically leads to a stable estimator, however, a misspecified one can result in unreliable conclusions.  Compared to linear models, nonparametric ones exhibit a higher degree of flexibility and have received recent attention in the functional context. We refer to \cite{ferraty2006nonparametric} and \cite{ling2018nonparametric} for general discussions on this topic. Various proposals  adapted nonparametric regression in the finite dimensional case to functional data, many of which are based on kernels that require distances between pairs of explanatory variables.  For functional data, natural measures of proximity involve notions of shape as those given by the similarities of derivatives or FPC scores \citep{shang2016bayesian}.  This suggests the use of semi-metrics as distance measures for many kernel-based functional regression estimators,  including  functional local-constant/linear estimators \citep{ferraty2002functional, baillo2009local,barrientos2010locally,berlinet2011local}, functional k-nearest neighbours \citep{burba2009k,kudraszow2013uniform, kara2017data}, functional recursive kernel estimator \citep{amiri2014recursive}, and Reproducing Kernel Hilbert Space (RKHS) methods \citep{preda2007regression, avery2014rkhs}. To avoid having to choose a semi-metric if several are available, \cite{ferraty2009additive} and \cite{febrero2013generalized}
suggested to combine kernel estimators constructed with different semi-metrics.

It is well known that in finite dimensional settings nonparametric estimators suffer from the ``curse of dimensionality"  due to the sparseness of data in high-dimensional spaces. Similarly, the performance of kernel estimators  for functional data is affected by sparsity in the infinite dimensional space, making it difficult to obtain a good estimate of the bandwidth \citep{geenens2011curse, mas2012lower}.  
When the distances between training
points and their closest neighbours require the use of large bandwidths the
resulting oversmoothing generally produces poor predictions for future observations. 
%When a data point has very few close neighbors, a small bandwidth may not contain enough neighbors to make  stable predictions, and a large one risks choosing some neighbors that do not share much similarity with that point.  
Exceptions derived from (generalized) additive models  avoid this issue by imposing some structure on the target function.   For example, \cite{muller2008functional} introduced a functional additive model with each component associated to a FPC score; 
\cite{muller2013continuously} and \cite{mclean2014functional}  studied  a continuous additive model $F(X) = \mu + \int G(X(t), t) dt$ and its generalized version  where the additivity for both models occurs in the time domain. Their methods use splines to fit the smooth regression surface $G$ with different regularization strategies. In addition to the above, \cite{gregorutti2015grouped} and \cite{moller2016random} used random forests 
with features extracted from functional variables  defined either as 
projections onto a functional basis (wavelet or FPC), 
or as averages of function values over randomly chosen intervals.

Beyond fully linear and nonparametric models, there have been  several developments towards semi-parametric regression for functional data, particularly in the direction of index-based models. \citet{lingsemiparametric} summarized advances in this field in a recent survey. \cite{ferraty2003modele} proposed a functional single index model  $F(X) =  r \left( \langle X , \beta \rangle \right)$, where $r: \mathbb{R} \rightarrow \mathbb{R}$ is an unknown smooth function and $\beta \in L^2(\mathcal{I})$ is the functional regression coefficient.   The computational and theoretical aspects of the proposal in \cite{ferraty2003modele} were further explored by \cite{ait2008cross}, \cite{ferraty2011estimation}, \cite{jiang2011functional}, and \cite{goia2015partitioned}.

To estimate more complex  regression functions,  some  proposals studied  multi-index models: 
\begin{equation} \label{eq:multi}
	F(X) = r(\langle X, \alpha_1 \rangle, ..., \langle X, \alpha_p \rangle),
\end{equation}
where $p$ is the number of indices and $r: \mathbb{R}^p \rightarrow \mathbb{R}$ is an unknown smooth function.  \cite{amato2006dimension} proposed an extension of 
 minimum average variance estimation to functional data. Their method requires a $p$-dimensional kernel smoother to estimate $r$, which depends on a computationally expensive procedure to select $p$ bandwidths and needs a large sample size to obtain a stable estimator when $p$ is large.  Other proposals were derived from sliced inverse regression \citep{li1991sliced}, including functional sliced inverse regression \citep{ferre2003functional}, functional k-means inverse regression \citep{wang2014functional}, and functional sliced average variance estimation  \citep{lian2014series}. These methods avoid $p$-dimensional smoothing but require  more 
restrictive assumptions on the distribution of $X$. 

Another approach uses a finite sums of single index terms: 
\begin{equation} \label{eq:index_approx}
	\hat{F}(X) = \hat{r}_1(\langle X, \hat{\alpha}_1 \rangle) + ... +  \hat{r}_p(\langle X, \hat{\alpha}_p \rangle). 
\end{equation}
This additive approximation is attractive since each component can be efficiently computed using a single dimensional smoother.  \cite{chen2011single} and \cite{ferraty2013functional} proposed iterative algorithms that compute the $\hat{r}_j$'s sequentially with kernel estimators, fitting one per iteration while keeping the previous ones fixed.   For non-Gaussian responses,  \cite{james2005functional} considered a generalized functional regression method with a known link function and fitted the $\hat{r}_j$'s  with spline smoothers.

In this article, we propose a new way to estimate the regression function using a boosting algorithm. We assume a nonparametric $F$ and approximate it with an additive multi-index estimator.  There exist other proposals that developed boosting-type algorithms for functional data, but they either apply to tasks different from ours (e.g. identifying the most predictive design points),  or do not consider multi-index estimators \citep{ferraty2010most, tutz2010feature, fan2015functional, ferraty2009additive,chen2011single, ferraty2013functional,greven2017general}. Compared to these proposals, our approach  fits more ``closely" into a boosting framework by minimizing a loss function with a gradient descent-like procedure and applying  regularization via shrinkage and early stopping.

Like gradient boosting, our algorithm constructs a regression estimator using a linear combination of ``base learners" 
 \begin{equation} \label{eq:index_multi}
	\hat{F}(X) = \hat{r}_1(\langle X, \hat{\beta}_{1,1} \rangle ,...,\langle X, \hat{\beta}_{1,K} \rangle ) + ... + \hat{r}_T(\langle X, \hat{\beta}_{T,1} \rangle ,...,\langle X, \hat{\beta}_{T,K} \rangle), 
\end{equation}
where each of the $T$ components $\hat{r}_j(\langle X, \hat{\beta}_{j,1} \rangle ,...,\langle X, \hat{\beta}_{j,K} \rangle )$ is characterized by $K$ indices.  
We fit each component with a tree-type base learner, which we call a ``functional multi-index tree", where the projection directions $\hat{\beta}_{j,1},..., \hat{\beta}_{j,K}$ and  $\hat{r}_j: \mathbb{R}^K \rightarrow \mathbb{R}$ are calculated  at the $j$-th boosting iteration. 

The base learners above have certain advantages over other possible choices one may 
consider. Specifically,
we do not need to rely on a specific parametric model, which may not work well if the 
data does not follow the chosen model. Non-tree non-parametric learners usually 
rely on pairwise distances or dissimilarities and require the carefully selection of 
bandwidths, which makes them notably computationally more costly than 
\eqref{eq:index_multi} above. Moreover, compared with the additive
structure of \eqref{eq:index_approx}, our approach 
enables modelling possible interactions between the indices, and thus is expected to work well for a much wider range of regression functions. 

The rest of this paper is organized as follows. 
In Section 2, we introduce the functional regression model and our boosting algorithm, including identifiability conditions for functional multi-index trees, and two algorithms to compute them.  Section 3 reports the results of a simulation study comparing our method with existing alternatives.     We discuss numerical results for a few representative settings and provide the complete set of  results in the Appendix. A case study is presented in Section~4, and Section~5 summarizes our findings and discusses directions for future work.

\section{Methodology}

Consider a data set  $(x_i, y_i), ..., (x_n, y_n)$ with  i.i.d. realizations of the pair $(X, Y)$, where $X$ are predictor variables in some space $\mathcal{X}$,  and $Y$ is a scalar response.  We are interested in estimating a function  $F: \mathcal{X} \rightarrow Y$ in order to make predictions for future observations of  $X$. 
  Following \cite{friedman2001greedy}, we define the target function $F$ as the minimizer of 
\begin{equation} \label{eq:obj}
	F \, = \argmin_G E_{Y,X} L(Y, G(X))
\end{equation}
over a class of functions ${\cal G}$, where $L$ is a pre-specified loss function such as the squared loss $L(a,b) = (a-b)^2$. Gradient boosting \citep{friedman2001greedy} is a method to estimate $F$ in \eqref{eq:obj} by constructing an approximate minimizer of the empirical risk: 
   \begin{equation} \label{eq:obj2}
 \argmin_{F} \sum_{i = 1}^n L (y_i, F(x_i)).  
 \end{equation}
 We view the objective in \eqref{eq:obj2}  as a function of the vector $(F(x_1), ..., F(x_n))^T$.  Similar to the  gradient descent algorithm that sequentially adds to the current point the negative gradient vector of the objective function, gradient boosting 
 adds to the current function estimate an approximation to the negative gradient vector. At the $t$-th iteration, the negative gradient vector $(u_{t,1},..., u_{t,n})^T$ is computed  at the point obtained from the previous iteration $(\hat{F}_{t-1}(x_1), ..., \hat{F}_{t-1}(x_n))^T$:   
  \begin{equation} \label{eq:u}
  	u_{t,i} = -\frac{\partial L(y_i, b)}{\partial b} \big\vert_{b = \hat{F}_{t-1}(x_i)},\  i = 1,...,n,
  \end{equation}
and is approximated using a base learner $\hat{h}_t: \mathcal{X} \rightarrow \mathbb{R}$ which we assume is characterized by parameters $\hat{\Theta}_t$ that are chosen to minimize  $$\sum_{i=1}^n(u_{t,i} - h_t(x_i; \Theta_t))^2.$$
%over members $h_t$ of a family of base learners $(\mathcal{H})$ and over $\Theta_t$ in the parameter space.  
We then calculate the step size
$$\hat{\alpha}_t = \argmin_{\alpha \in \mathbb{R}}  \sum_{i=1}^n L(y_i, \hat{F}_{t-1}(x_i) + \alpha \hat{h}_t(x_i; \hat{\Theta}_t))$$
and update $\hat{F}_t$ using 
a shrinkage
parameter $\gamma \in (0, 1)$ to control the learning rate:
\begin{equation}\label{eq:update}
\hat{F}_{t}(x) = \hat{F}_{t-1}(x) + 
 \gamma \bigl[ \hat{\alpha_t} \hat{h}_t(x;\hat{\Theta}_t ) \bigr] =
\hat{F}_{t-1}(x) + 
\gamma \hat{\alpha_t} \hat{h}_t(x;\hat{\Theta}_t ) \, .
\end{equation}
The inclusion of $\gamma$ is expected to reduce the impact of each base learner $\hat{h}_t$ on  $\hat{F}_t$  by approximating more finely the search
path, and as a result, improve the prediction accuracy \citep{telgarsky2013margins}.  This simple and effective strategy can be viewed as a form of regularization.  \cite{friedman2001greedy} showed empirically that lower prediction errors can be achieved when $\gamma \leq 0.1$. 

Similar to what the gradient descent algorithm does, gradient boosting minimizes the training loss in a greedy fashion and may eventually overfit  the training data at the expense of degraded generalization performance. To avoid this issue, boosting algorithms generally use an ``early stopping" strategy that stops training  when the performance on the validation set starts to deteriorate.  We define an early stopping rule based on the  performance on a validation set ($\mathcal{V}$) that follows the same model as the training set. Usually, the validation set is randomly selected from  all available data and set aside from the training set.  The  early stopping time is defined as the iteration that achieves the lowest loss  on  $\mathcal{V}$: 
\begin{equation}  \label{eq:early}
	T_{\text{stop}} = \argmin_{t = 1,...,T_{\text{max}}}\sum_{i \in \mathcal{V}} L(y_i, \hat{F}_{t}(x_i)), 
\end{equation}
where $T_{\text{max}}$ is the maximum number of iterations allowed in our algorithm.  After  training completes, the final boosting estimator is given by 
 $$\hat{F}_{T{\text{stop}}}(x) = \hat{F}_{0}(x) + \sum_{t=1}^{T_{\text{stop}}}\gamma \alpha_t \hat{h}_t(x; \hat{\Theta}_t),$$
 where the initial function estimate is generally a constant defined as 
 $$\hat{F}_0(x) = \argmin_{a\in \mathbb{R}} \sum_{i=1}^n L(y_i, a). $$
 
 Core to the performance of a boosting algorithm is the choice of base learners.  In the functional context,  for $X$ in a Hilbert space $\mathcal{L}$ (e.g. $L^2(\mathcal{I})$ the space of square-integrable functions) we will introduce a flexible class of base learners which we call ``functional multi-index trees". They have the advantage of being capable of fitting  functions of $X$ that involve multiple projections and their possible interactions.  Furthermore, they are scalable to large sample sizes and  can easily incorporate additional real-valued explanatory variables. 

We can  adapt the proofs in \cite{zhang2005boosting} (Theorem 4.1 and Appendix A.3) to establish the convergence of our estimator when $\gamma = 1$.  Specifically, for any given training set  $(x_i, y_i), ..., (x_n, y_n)$, let $\hat{A}_n(G) =  \sum_{i = 1}^n L (y_i, G(x_i))$ for $G \in  \mathcal{G} = \left\{ \sum_{t = 1}^T \alpha_t  h_t(\cdot; \Theta_t),  T \in \mathbb{N},  \alpha_t \in \mathbb{R} \right\}$, and $h_t$ belongs to the class of functional multi-index trees. Then, 
$\lim_{t\rightarrow \infty} \hat{A}_n(\hat{F}_t) = \inf_{G\in \mathcal{G}} \hat{A}_n(G)$. 
These results also hold at the population level with $A(G) = E_{Y,X}L(Y, G(X))$ and $\hat{F}_t$ constructed using $\hat{\alpha}_t$ that satisfies $\hat{\alpha}_t = \argmin_{\alpha \in \mathbb{R}} E_{Y,X} L(Y, \hat{F}_{t-1}(X) + \alpha \hat{h}_t (X; \hat{\Theta}_t))$. 
Furthermore, \cite{zhang2005boosting} showed the convergence and consistency of boosting estimators with restricted step sizes and early stopping times linked to step sizes. While the details of their regularization strategies are different from ours,  their findings provide support for the use of shrinkage and early stopping considered by our method. 

\subsection{Functional multi-index trees} \label{sec:tree} 
 When the explanatory variables are vectors in $\mathbb{R}^d$ for some $d>1$, decision trees are commonly used as base learners for gradient boosting, which select variables to make the best splits. Similarly for $X \in \mathcal{L}$, we ``select" the optimal indices and use them to  define the splits that partition the data.

Let  $\beta_1, , ..., \beta_K \in \mathcal{L}$  be $K$ functions which we view as  projection directions. The inner-products  $\langle x, \beta_1 \rangle, ..., \langle x, \beta_K \rangle$ are corresponding indices projecting $x \in \mathcal{L}$ onto these directions. 
  At the $t$-th iteration, we compute negative gradients $u_{t,i}$ as in \eqref{eq:u} and
define a functional multi-index tree $h(\langle \cdot , \beta_1 \rangle, ..., \langle \cdot , \beta_K \rangle)$ as the solution to 
 \begin{equation} \label{eq:est}
\argmin_{h, \beta_1,...,\beta_K} \sum_{i=1}^n \left(u_{t,i} - h \left(\langle x_i, \beta_1 \rangle, ..., \langle x_i, \beta_K \rangle \right) \right)^2,
\end{equation}
where $h: \mathbb{R}^K \rightarrow \mathbb{R}$ is a decision tree.  Below we discuss two strategies to compute functional multi-index trees.
We refer to the resulting trees as Type A and Type B trees.   
%: one chooses the optimal set of $K$ indices over the entire tree, and the other finds a best single index at each split.  We refer to the resulting trees as Type A and Type B trees, respectively. 

%Based on characteristics of the data and computation efficiency, we will make recommendations
%We also  on which one to use  
%It is easy to see that the solution to \ref{eq:est} is not unique if $\beta_1,...,\beta_K$ are unconstrained. For example, we can write  $g(\langle \cdot , \beta_1\rangle, ..., \langle  \cdot , \beta_K\rangle) = g(\langle \cdot , c_1\beta_1\rangle, ..., \langle \cdot , c_K\beta_K\rangle)$ for any non-zero real values $c_1, ..., c_k$. 

 %There are several methods to make a semiparametric model identifiable, among which we suggest one that 

\subsubsection{Type A trees}
\label{subsec:typeA}
The Type A tree takes a two-level approach where we  estimate 
$h$ at the inner level and search for the optimal $\hat{\beta}_1,..., \hat{\beta}_K$ at the outer level. It is easy to see this two-level structure if we re-state  \eqref{eq:est} as 
	\begin{equation}   \label{eq:exp}
	\argmin_{\mathbf{\beta}_1,...,\mathbf{\beta}_K} \left\{\argmin_{h} \sum_{i=1}^n \left(u_{t,i} - h \left(\langle x_i, \beta_1 \rangle, ..., \langle x_i, \beta_K \rangle \right) \right)^2 \right\}. 
\end{equation}
Given $\beta_1,..., \beta_K$, the solution to the inner optimization can be approximated by fitting  a decision tree with the CART algorithm \citep{breiman1984} using  $\langle x_i, \beta_1 \rangle$, ...,  $\langle x_i, \beta_K \rangle$, $i = 1,..., n$ as predictors. At the outer level, we find $\beta_1$,..., $\beta_K$  that minimize 
\begin{equation} \label{eq:est2}
	\sum_{i=1}^n (u_{t,i} - \hat{h} (\langle x_i, \beta_1\rangle,..., \langle x_i, \beta_K\rangle ))^2. 
\end{equation}
 If $\beta_1,...,\beta_K$ are unconstrained,   the solution to \eqref{eq:exp} is not unique,
which may cause convergence problems for optimization algorithms. 
 For  example, for any decision tree $h$  and non-zero real values $b_1, ..., b_k$,   $$h(\langle \cdot , \beta_1\rangle, ..., \langle  \cdot , \beta_K\rangle) = \tilde{h}(\langle \cdot , b_1\beta_1\rangle, ..., \langle \cdot , b_K\beta_K\rangle),$$  where the value at which the $j$-th input of $h$ is split is multiplied by $b_j$ in $\tilde{h}$.   
To avoid this issue, we introduce conditions  under which $h$, $\beta_1$, ..., and $\beta_K$ are identifiable up to sign changes of each $\beta_j$.
Similar conditions have been considered previously in the literature, see e.g. \cite{ait2008cross} and \cite{ferraty2013functional}.

%Inner level (estimating $h$): For  $\mathbf{z}_i= (z_{i,1}, ..., z_{i,K}) \in \mathbb{R}^K$ where $z_{i,j} = \langle x_i, \beta_j \rangle  $,  a  tree $h$ partitions the space of all possible values of $\mathbf{z}$ into disjoint regions $R_j, j = 1,...,J$, represented by the terminal nodes of the tree.  Given $\mathbf{z}_1, ..., \mathbf{z}_n,$ $\hat{h}$ is the solution that minimizes the empirical risk $$\sum_{i=1}^n \left(u_i - \sum_{j = 1}^J a_jI(\mathbf{z}_i \in R_j) \right)^2$$
%over all possible values of $a_j$ and partitions of regions $R_j$, where $I$ is an indicator function that has value 1 if its argument is true. This implies that $\hat{h}$ can be viewed as an implicit function of $\mathbf{z}_i$, and further of $\beta_1$,..., $\beta_K$ given $x_i, ..., x_n$. 

%Outer level (estimating $\beta$): We search for $\beta_1$,..., $\beta_K$  that minimizes 

\begin{condition} \label{condition:iden1}
$\lVert\beta_j\rVert = 1$, for $j = 1,...,K.$
\end{condition}

\begin{condition}  \label{condition:iden2}
Each index is chosen by at least one of the splits of a decision tree $h: \mathbb{R}^K \rightarrow \mathbb{R}$.  For any set of indices $J \subset\{1,...,K\}$, there exist a $\delta>0$ and  $x_0 \in B(x_0, \delta) =  \{x \vert x \in \mathcal{L}, \lVert x - x_0\rVert\leq \delta\}$, 
such that for $x \in B(x_0, \delta)$, 
\begin{equation}\label{eq:proof}
	h(\langle  L_1(x), \beta_1 \rangle ..., \langle L_j(x), \beta_j \rangle ..., \langle L_K(x), \beta_K \rangle )
\end{equation}
is a non-constant function of $x$, where  $L_j(x) = x$ if $j \in J$  and else $L_j(x) = x_0$,  for $j = 1,..., K$. 
\end{condition}

The following result shows that these conditions are sufficient to generate a unique solution to \eqref{eq:exp} up to sign changes of each $\beta_j$.  The proof of \Cref{thm:1} is included in the Appendix.

\begin{thm} \label{thm:1}
Suppose that Conditions \ref{condition:iden1} and \ref{condition:iden2} hold, then $\beta_1$,..., $\beta_K$  are identifiable up to sign changes of each $\beta_j$.  That means for any decision trees $h: \mathbb{R}^K \rightarrow \mathbb{R}$ and $\tilde{h}: \mathbb{R}^K \rightarrow \mathbb{R}$ if 
\begin{equation}\label{eq:thm}
	h(\langle x, \beta_1\rangle ...,  \langle x, \beta_K \rangle)= \tilde{h}(\langle x, \eta_1 \rangle, ..., \langle x, \eta_K \rangle)
\end{equation}
hold for all $x \in \mathcal{L}$, 
then $\{\beta_1,..., \beta_K \} = \{(-1)^{l_1}\eta_1,..., (-1)^{l_K}\eta_K \}$ for some $l_1,...,l_K \in \{0,1\}$, and $h = \tilde{h}$. 
\end{thm}
 A convenient approach  to find the solution of \eqref{eq:exp} is to approximate the optimal directions using a flexible basis (e.g. splines).  Let $\{\psi_1$,..., $\psi_s\}$ be an orthonormal set in 
$\mathcal{L}$  and write $\beta_j =  \sum_{l = 1}^sc_{j,l} \psi_l$.   Then \Cref{condition:iden1}  becomes $\rVert \mathbf{c}_j \rVert_2 = 1$  where  $ \rVert \cdot \rVert_2$ is the Euclidean norm and  $ \mathbf{c}_j = (c_{j,1}, ...,c_{j,s})^T$  for $j = 1,..., K$. \Cref{condition:iden2} generally holds for functional multi-index trees except for very special situations,  such as one of the indices being equal for all individuals. 

With a slight abuse of notation, we define  $\langle \cdot  , \boldsymbol{\psi} \rangle  = (\langle \cdot , \psi_1 \rangle ,..., \langle \cdot , \psi_s \rangle)^T$ and write the objective in \eqref{eq:exp} as 
\begin{equation}\label{eq:est3}
	\sum_{i=1}^n (u_{t,i} - h( \mathbf{c}_1^T \langle x_i , \boldsymbol{\psi}\rangle, ...,   \mathbf{c}_K^T\langle  x_i , \boldsymbol{\psi}\rangle))^2. 
\end{equation}
To further simplify the computation to minimize \eqref{eq:est3} under the condition $\rVert \mathbf{c}_j \rVert_2 = 1$ for $j = 1, ..., K$, we represent $\mathbf{c}_j$ in a spherical coordinate system in order to obtain an optimization problem with simple box constraints.  

Referring to \cite{blumenson1960derivation},  Cartesian coordinates $\mathbf{c}_j$ and  spherical coordinates $(r_j, \theta_{j,1},..., \theta_{j,s-1})$ are connected through the following equations: 
\begin{align*}
&r_j = \sqrt{\sum_{l=1}^sc_{j,l}^2} \\
&c_{j,1} = 	r_j \text{cos}(\theta_{j,1}) \\
&c_{j,l} = 	r_j \text{cos}(\theta_{j,l}) \prod_{k=1}^{l-1} \text{sin}(\theta_{j,k}), \ l = 2,..., {s-1} \\ 
&\vdots \\
& c_{j,s} = r_j \prod_{k=1}^{s-1}\text{sin}(\theta_{j,k}), 
\end{align*}
where $r_j \in [0, \infty)$, $\theta_{j,1} \in [-\pi,\pi)$, and $\theta_{j,l} \in [0,\pi]$  for   $l = 2,..., {s-1}$.  As a result of this connection, we can transform $\mathbf{c}_j$ to  $(r_j, \theta_{j,1},..., \theta_{j,s-1})$ and further reduce \Cref{condition:iden1} to  $r_j = 1$. In addition, we restrict $\theta_{j,1} \in [-\pi/2, \pi/2)$, which is equivalent to $c_{j,1} > 0$. This restriction ensures that \eqref{eq:est3} has a single unique solution as opposed to multiple solutions that are unique up to sign changes. 

Finally, we treat \eqref{eq:est3} as a function of  $\boldsymbol{\theta}_j = (\theta_{j,1},..., \theta_{j,s})^T$ and  minimize 
it under conditions 
\begin{equation}\label{eq:box}
\theta_{j,1} \in[-\pi/2, \pi/2),  \ \theta_{j,2},...,\theta_{j,s-1} \in [0,\pi]. 
\end{equation}
This can be achieved by applying generic gradient-free optimization algorithms that allow box constraints.  Note that the objective function \eqref{eq:est3} may not be convex and therefore multiple starting points are recommended when preforming the optimization (see \Cref{sec:imp} for details).

\subsubsection{Type B trees}
For this approach, instead of building a tree for each possible set of 
directions, and then optimizing the fit over the directions, we 
find optimal directions as the tree is being built. 
Due to the 
recursive binary search strategy behind the CART algorithm \citep{breiman1984}, 
this is equivalent to selecting an optimal direction (from the random set of 
candidates) at each split. 
The number of indices of Type B trees  is determined by the maximum depth of the tree. 
Thus, 
at each boosting iteration $t$, 
we find the optimal index for one split at a time.  At the first split 
we find 
\begin{equation} \label{eq:exp3}
	\argmin_{\beta} \left \{\argmin_{g \in G_1} \sum_{i=1}^n (u_{t,i} - g \left (\langle x_i, \beta \rangle \right)^2\right\},
\end{equation}
where $G_1$ denotes the class of single splits, also called decision stumps.  

For any direction $\beta$, the tree $g$ partitions the data into two regions (or nodes) where the dividing threshold is selected from  $\langle x_1, \beta \rangle, ..., \langle x_n, \beta \rangle$ to minimize the squared prediction error. 
%The predicted value of each data point is the average  $u_{t,i}$'s in the region that the data point belongs to.
If the tree in \eqref{eq:est} has depth larger than 1, 
we solve \eqref{eq:exp3} for each of the ``children'' nodes, where the 
sum now is over the points in each region. This process is repeated 
recusively until we reach the specified maximum tree depth. 

We can view \eqref{eq:exp3} as fitting a Type A tree with $K = 1$ and depth $d = 1$, and apply the algorithm in \Cref{subsec:typeA} at every split. However, this can be computationally costly when a Type B tree contains many splits.  There is also the concern of performing optimization at nodes that contain very few examples, for instance, nodes at deep levels of the tree, which may result in unstable estimation of $\beta$ or even  convergence failure of the optimization algorithm. To avoid these issues,  we suggest an alternative approach to approximate the solution of \eqref{eq:exp3}. 

Rather than solving \eqref{eq:exp3}  with respect to all possible $\beta$s at each split, we select  $\beta$ from a  pool of randomly generated candidates.  As we did for Type A trees, we expand $\beta$ in the space of an orthonormal basis  $\{\psi_1$,..., $\psi_s\}$ and let $\beta = \sum_{l = 1}^sc_{0,l} \psi_l$.   We randomly generate a large pool of candidate $\mathbf{c}_0$ vectors denoted as $\mathcal{C}^{(t)} = \{\mathbf{\tilde{c}}_{1}^{(t)},...,\mathbf{\tilde{c}}_{P}^{(t)}\}$, where each $\mathbf{\tilde{c}}_{j}^{(t)} \in \mathcal{C}^{(t)}$ is an independently sampled vector $\mathbf{\tilde{c}}_{j}^{(t)} = (\tilde{c}_{j,1}^{(t)}, ..., \tilde{c}_{j,s}^{(t)})^T$ that satisfies $\lVert\mathbf{\tilde{c}}_{j}^{(t)}\rVert_2 = 1$ and  $\tilde{c}_{j,1}^{(t)} >0$. This ensures the candidate $\beta$s satisfy identifiability conditions $\lVert \tilde{\beta}_j \rVert  = 1$ where $\tilde{\beta}_j = \sum_{l=1}^s\tilde{c}^{(t)}_{j,l}\psi_l$ for $j = 1,...,P$. 
 
 At each split, the coefficient vector of the projection direction is selected from $\mathcal{C}^{(t)}$. This corresponds to applying the CART algorithm \citep{breiman1984} to find a decision tree $h: \mathbb{R}^{P} \rightarrow \mathbb{R}$ that minimizes the squared error: 
\begin{equation} \label{eq:approx}
	\argmin_{h} \sum_{i=1}^n \left(u_{t,i} - h\left( (\mathbf{\tilde{c}}_{1}^{(t)})^T \left \langle x_i, \boldsymbol{\psi} \right \rangle, ..., (\mathbf{\tilde{c}}_{P}^{(t)})^T \left \langle x_i,  \boldsymbol{\psi} \right \rangle  \right) \right)^2,
\end{equation}
where  $ (\mathbf{\tilde{c}}_{1}^{(t)})^T \left \langle x_i, \boldsymbol{\psi} \right \rangle, ..., (\mathbf{\tilde{c}}_{P}^{(t)})^T \left \langle x_i,  \boldsymbol{\psi} \right \rangle $ are the explanatory variables and $u_{t,i}$ is the response for the $i$-th individual. In order to reduce overfitting, we use different  pools $C^{(t)}$ randomly chosen at each boosting iteration. This is expected to introduce randomness to the algorithm and allow more directions to be considered.

\subsubsection{Comparison of Type A and Type B trees}

Type A and Type B trees differ in the way they find an approximate solution to \eqref{eq:est}. 
In particular, with Type A trees 
we fit a tree for each set of directions
$\beta_j$, $1 \le j \le K$, and then optimize (numerically) 
the resulting fit over the directions, while for Type B trees we find optimal directions 
as we construct the tree. Furthermore, these directions 
are chosen from a candidate pool of randomly generated $\beta$'s. 

When feasible, Type A trees are expected to provide a better approximate solution to 
\eqref{eq:est} than Type B trees, given that for the latter we only use a set of
randomly generated candidate indices. However, it can be computationally prohibitive to fit Type A trees with a moderate or large value of $K$, which may be needed to obtain reliable 
predictions for complex regression functions. In these cases Type B trees provide a flexible structure that is much faster to compute. 
In summary, if the target function is believed to be complex we suggest using Type B trees
to take advantage of their fast computation and flexibility.   On the other hand, when the target function is simple,  parsimonious base learners (e.g. Type A trees with few indices) are often preferred to prevent overfitting.  

Note that both Type A and B trees can easily include real-valued explanatory variables $\mathbf{v}_{i} \in \mathbb{R}^q$ by replacing $h(\langle x_i, \beta_1\rangle,..., \langle x_i, \beta_K\rangle)$ in \eqref{eq:est} with $h(\langle x_i, \beta_1\rangle,..., \langle x_i, \beta_K \rangle, \mathbf{v}_i)$.  More specifically,  \eqref{eq:est3} becomes 
\begin{equation}
	\sum_{i=1}^n (u_{t,i} - h( \mathbf{c}_1^T \langle x_i , \boldsymbol{\psi}\rangle, ...,   \mathbf{c}_K^T\langle  x_i , \boldsymbol{\psi}\rangle, \mathbf{v}_i))^2, 
\end{equation}
for Type A trees, and  \eqref{eq:approx} becomes 
\begin{equation} 	\argmin_{h} \sum_{i=1}^n \left(u_{t,i} - h\left( (\mathbf{\tilde{c}}_{1}^{(t)})^T \left \langle x_i, \boldsymbol{\psi} \right \rangle, ..., (\mathbf{\tilde{c}}_{P}^{(t)})^T \left \langle x_i,  \boldsymbol{\psi} \right \rangle,    \mathbf{v}_i \right) \right)^2
\end{equation}
for Type B rrees. 
This extension allows our method to fit 
partial-functional regression models with mixed-type predictors.  In \Cref{sec:real}, we will introduce an example that illustrates the usage of our method in a partial-functional setting.

\section{Simulation} \label{Sec:sim}
To assess the numerical performance of our proposed method, we conducted a simulation study comparing it with alternative  functional regression methods in the literature.

We generated data sets $D = \{(x_i, y_i), i = 1,..., N\}$, consisting of a predictor $x_i\in L^2(\mathcal{I})$ and a scalar response $y_i$ that follow the model: 
 \begin{equation} \label{eq:gen}
y_i = r(x_i) + \rho \epsilon_i,
\end{equation}
where the errors $\epsilon_i$ are i.i.d $N(0,1)$, $r$ is the regression function,  and  $\rho > 0$ is a constant that controls the signal-to-noise ratio (SNR): 
$$\text{SNR} = \frac{\text{Var}(r(X))}{\text{Var}(\rho\epsilon)}.$$

\begin{figure}
  \centering
 	\includegraphics[scale=0.45]{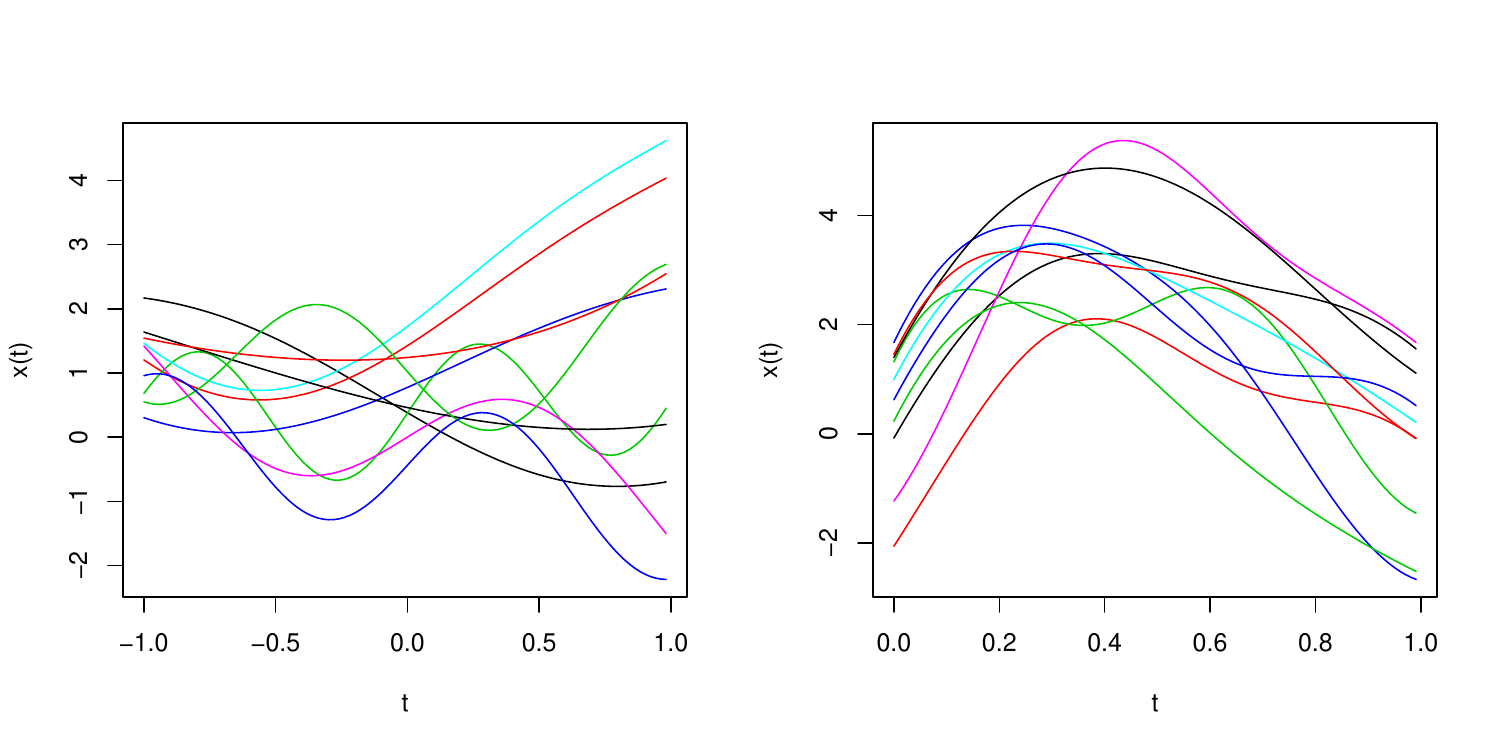}
 	\caption{Ten random samples generated from Model 1 (left panel) and Model 2 (right panel). }
 	 \label{fig:x}
 \end{figure}

To sample the functional predictors $x_i$, we considered two models adopted from  \cite{ferraty2013functional} and \cite{boente2021robust} respectively: 

\begin{itemize}
	\item[-] Model 1 $(M_1)$: 
	$$x_i(t) = a_i + b_i t^2 + c_i \, \text{exp}(t) + \text{sin}(d_it),$$
  where $t \in [-1,1]$, $a_i, b_i \sim \mathcal{U}(0,1)$, $c_i \sim \mathcal{U}(-1,1)$, and $d_i \sim \mathcal{U}(-2\pi, 2\pi)$; and   
  \item[-] Model 2 $(M_2)$: 
        $$x_i(t) = \mu(t) + \sum_{p=1}^4 \sqrt{\lambda_j}\xi_{ij}\phi_j(t),$$
where $t \in [0,1]$, $\mu(t) = 2\,  \text{sin}\, (t\pi) \, \text{exp}(1-t)$, $\lambda_1 = 0.8, \lambda_2 = 0.3, \lambda_3 = 0.2$, and $\lambda_4 = 0.1$,  $\xi_{ij}\sim N(0,1) $,  and $\phi_j$'s are the first four eigenfunctions of the ``Mattern'' covariance function $\gamma(s,t)$ with parameters $\rho = 3, \sigma = 1, \nu = 1/3$: 
 $$\gamma(s,t) = C\left(\frac{\sqrt{2\nu}|s-t|}{\rho}\right), \ C(u) = \frac{\sigma^2 2^{1-\nu}}{\Gamma(\nu)} u^{\nu} K_{\nu}(u),$$
  where $\Gamma(.)$ is the Gamma function and $K_{\nu}$ is the modified Bessel function of the second kind.
\end{itemize}
For each subject $i$, we evaluated $x_i$ on a dense  grid $t_1,..., t_{100}$  in $\mathcal{I} = [-1,1]$ for $M_1$  and $\mathcal{I} = [0,1]$ for $M_2$.  \Cref{fig:x} shows an example of 10 randomly sampled  $x_i$'s generated from  $M_1$ (left panel) and $M_2$ (right panel).   

We considered five regression functions:
\begin{itemize}

\item[- ]  $r_1(X) = \left(\int_{\mathcal{I}} (X(t) - \mu(t)) (\phi_1(t) + \phi_2(t))dt \right)^{1/3},$ where the first two FPCs ($\phi_1$ and $\phi_2$) and mean ($\mu$) for $M_1$ were calculated using a  large sample of  $x_i, i = 1,..., 3000$, while  for $M_2$ their true values were used, 
\item[- ]  $r_2(X) = 5 \, \text{exp}\left (- \frac{1}{2} \lvert \int_{\mathcal{I}} x(t)\log(|x(t)|)dt \rvert \right),$
\item[- ] 
$r_3(X) = 5 \, \text{logistic}\left(2\int_{\mathcal{I}}X(t)^2 \text{sin}(2\pi t) dt \right),$ where  $\text{logistic}(u) = 1/(1+ \text{exp}(-u))$, and
\item[- ] 
$r_4(X) = 5 \left( \sqrt{\lvert\int_{\mathcal{I}_1} \text{cos}(2\pi t^2) X(t) dt \rvert} + \sqrt{\lvert \int_{\mathcal{I}_2} \text{sin}(X(t)) dt \rvert} \right), $ where  $\mathcal{I}_1 = [-1,0]$ and $\mathcal{I}_2 = (0,1]$ for $M_1$, and  $\mathcal{I}_1 = [0,0.5]$ and $\mathcal{I}_2 = (0.5,1]$ for $M_2$. 
%\item[- ]  $r_5(X) =  \int_{\mathcal{I}} \left (\text{sin} \left(\frac{3}{2} \pi t \right) +  \text{sin} \left(\frac{1}{2} \pi t \right)\right)X(t)dt,$
\end{itemize}

These regression functions were selected to represent a single-index model $(r_1)$ and other nonlinear models ($r_2$, $r_3$, and $r_4$). To control the level of noise,  we considered SNR = 5 and 20  as high and low  noise levels and denoted them as $S_1$ and $S_2$,  For each combination of the predictor model ($M$), regression function ($r$), and noise level ($S$), we generated i.i.d. samples $(x_i, y_i),...,(x_N, y_N)$ of size $N = 1600$.  The dataset was randomly partitioned  into a training set, a validation set, and a test set of size 400, 200, and 1000 respectively.   In total, we simulated data from 16 settings, each defined as  a combination $(M, r, S)$ from the set $\{M_1, M_2\} \times  \{r_1,...,r_4\}\times \{S_1,S_2\}$.

\subsection{Implementation details} \label{sec:imp}
 For each setting, we used 100 independently generated datasets and compared the performance of the following estimators: 

\begin{itemize}
 \setlength\itemsep{0.1em}
  \item[- ]  \texttt{FLM1}: functional linear regression  with cubic B-splines  \citep{hastie1993statistical},
  \item[- ]  \texttt{FLM2}: functional linear regression with FPC scores \citep{cardot1999functional},
\item[- ]  \texttt{FAM}: functional additive models \citep{muller2008functional},
 \item[- ]  \texttt{FAME}: functional adaptive model estimation  \citep{james2005functional},  
\item[- ]  \texttt{FPPR}: functional projection pursuit regression \citep{ferraty2013functional},
\item[- ]  \texttt{FGAM}: functional generalized additive models \citep{mclean2014functional}, 
\item[- ]  \texttt{RFgroove}: random forests for grouped functional variable selection \citep{gregorutti2015grouped} 
\item[- ]  \texttt{FRF}: functional random forest with random intervals  \citep{gregorutti2015grouped} 
\item[- ]  \texttt{TFBoost(A,K)}: tree-based functional boosting with Type A trees,  where each tree is specified with  $K$ = 1, 2, or 3 indices, and 
\item[- ]   \texttt{TFBoost(B)}: tree-based functional boosting with  Type B trees.  
\end{itemize}

\texttt{FLM1} and \texttt{FLM2} are classical functional regression methods for linear models, and they use different bases to estimate the regression coefficients.   For \texttt{FLM1}, 
we used  cubic B-splines with 7  basis functions (3 evenly spaced interior knots), which was found to work generally well compared to using less or more interior knots.  We penalized the second derivative of the coefficient and selected the penalization parameter to minimize the   mean-squared-error (MSE) on the validation set. For \texttt{FLM2}, we used the top 4 FPCs required to explain 99\% of the variance. 

\texttt{FAM} constructs an additive estimator using FPC scores as predictors. We referred to its implementation in the \texttt{fdapace} package  \citep{rfdapace} and used the Gaussian kernel to fit each additive component. Same as what we did for \texttt{FLM2}, the top 4 FPCs were used for \texttt{FAM}.  

\texttt{FAME} estimates an extension of generalized additive models (GAM) to functional predictors. We used the code shared by the authors of \cite{james2005functional} and fitted the model using a Gaussian link function. \texttt{FPPR} follows the principle of projection pursuit regression and  assumes an additive decomposition with each component being a functional single index model (FSIM).  We implemented \texttt{FPPR}  using the code to fit a single additive component shared by the authors of \cite{ferraty2013functional} and built an estimator by adding multiple functional single index estimators.  Similarly to what was done with \texttt{FLM1}, for \texttt{FPPR} and \texttt{FAME} we used cubic B-splines with 7 functions (3 evenly spaced interior knots) as the basis. For both methods, we selected the number of additive components between 1 and 15 to minimize the MSE on the validation set. In our simulation settings, the performance of \texttt{FAME} and \texttt{FPPR} rarely improved beyond  15 additive components.  

The implementation of \texttt{FGAM} is available from the \texttt{refund} package in R \citep{rrefund}. The method fits a model of the form 
\begin{equation*}
F(X) = \mu + \int G(X(t), t) dt,
\end{equation*}
where $\mu\in \mathbb{R}$, and $G$ is estimated using bivariate B-splines with roughness penalties.  A restricted maximum likelihood approach is applied to select the parameter that penalizes the second order marginal differences of the basis (see the documentation of \texttt{fgam} in the \texttt{refund} package for details). We chose to use tensor-type bivariate cubic B-splines of dimensions 15 by 15. In all our settings, this choice ensured a stable fit and almost always resulted in the best performance compared to using other B-splines bases. 

The method in \texttt{RFgroove} was originally proposed for computing importance measures for groups of functional variables \citep{gregorutti2015grouped}. Although the concept of grouped variable importance  is not applicable to our case with a single functional predictor, we borrowed the idea to fit a random forest using projections on a functional basis 
as features.  The R package \texttt{RFgroove} \citep{RFgroove}
implements wavelet and FPC bases. 
However the discrete wavelet transform (DWT) requires data of length $2^J$ for some integer $J$, which was not applicable in our case, so we used the 
top FPC basis components that explain 99\% of the variance. 
We adapted the code to obtain predicted values and used 500 trees to construct the random forest with the default control parameters 
of the function \texttt{rpart} that fits the trees. 
%the package does not provide the predicted values for the response and only  returns the variable importance.  So we adapted the code and use the top FPC basis components selected to explain 99\% of the variance.
%we used the top 4 FPCs as the basis functions to explain  99\% of the variance. 

The \texttt{FRF} approach 
of \cite{moller2016random}
randomly partitions the domain of the functional variables
into intervals and uses the average of the function evaluations over these intervals as predictors in regression trees. Following \cite{moller2016random}, we used 500 regression trees to construct a random forest, sampled the lengths of the intervals from the exponential distribution, and selected the rate parameters from the values \{1, 5, 10, 15, 20, 25, 30\} in order to minimize the prediction error on the validation set.

 For the proposed methods, we implemented \texttt{TFBoost} in R,  including \texttt{TFBoost(A.K)} for positive integers $K$ and \texttt{TFBoost(B)}. We fitted decision trees using the \texttt{rpart} package \citep{rpart} and performed optimization of \texttt{TFBoost(A.K)}  using the \texttt{Nelder\_Mead} function from the \texttt{lme4} package \citep{lme4}.   The code implementing our method is publicly available online at \texttt{https://github.com/xmengju/TFBoost}.  To make a fair comparison with its alternatives, we used  cubic B-splines with 7 functions (3 evenly spaced interior knots) as the basis for \texttt{TFBoost} methods, same as what we did for \texttt{FLM1}, \texttt{FPPR}, and \texttt{FAME}. We then  orthonormalized the basis as required by Type A and Type B trees.

 We studied the performance  of \texttt{TFBoost(A.K)} with $K$ = 1,2, and 3; and \texttt{TFBoost(B)} using 200 random directions at each iteration. For each of these methods, the maximum depth ($d$) of the functional multi-index trees was fixed for all iterations. We experimented with  $d \in \{1,2,3,4\}$ and for each method selected the value of $d$ that achieved the lowest MSE on the validation set at the early stopping time.  We set the shrinkage parameter $\gamma$ to 0.05 and  the maximum number of iterations $T_{\texttt{max}}$  to 1000.  

As mentioned in \Cref{subsec:typeA},  the estimation of a Type A tree may result in a non-convex optimization problem.  To avoid suboptimal local minima, we fitted each Type A tree with multiple starting points.  More specifically,  we first uniformly sampled 30 points in the spherical coordinates system that satisfied  the box constrains in \eqref{eq:box} and  ran the Nelder-Mead algorithm for 10 steps using each of the 30 points as the starting point.  We then chose the five ending points with the lowest objective values and continued running the Nelder-Mead algorithm until convergence. 
Out of the five resulting estimates, we chose the one that minimized the objective function. 

The results for all methods under consideration were evaluated in terms of the MSE on the test set ($\mathcal{T}$): 
\begin{equation}\label{eq:MSE}
	\text{MSE} = \frac{1}{\vert\mathcal{T}\vert} \sum_{i \in \mathcal{T}} (\hat{F}(x_i) - y_i)^2.
\end{equation}
For \texttt{TFBoost}, the estimate $\hat{F}$ was reported at the early stopping time.

\subsection{Results}
For each combination of the regression function $(r_1,...,r_4)$ and the predictor model ($M_1, M_2$), the results for the high noise ($S_1$)  settings look very similar as those for the low noise ($S_2$) settings.  In $S_2$ settings, the differences between  estimators are more pronounced, showing a larger advantage of \texttt{TFBoost(A.K)} and \texttt{TFBoost(B)} over the others.   We report here the results for $S_1$ settings and provide those for $S_2$ in the Appendix.

\Cref{fig:sim1,fig:sim2,fig:sim3,fig:sim4} show the MSEs on the test sets  for $r_1$ to $r_4$ respectively. Since \texttt{FLM1}, \texttt{FLM2}, \texttt{FAM}, and  \texttt{RFgroove} tend to produce very large errors,  to be able to visualize the differences among the other methods, we excluded them from the figures and reported the summary statistics of their MSEs in the Appendix.   As expected, the linear estimators (\texttt{FLM1} and \texttt{FLM2})  do not work well  since $r_1$ to $r_4$ are nonlinear functions.  \texttt{FAM} and \texttt{RFgroove}  also perform poorly, likely due to using FPCs as the projection direction to construct features, making them less flexible compared with using random features (like those generated by \texttt{FRF}) or estimating the projection directions based on the data. 
 
\begin{figure}[htp]
     \centering
     \begin{subfigure}[b]{0.49\textwidth}
         \centering
         \includegraphics[width=1.05\textwidth, height = 4.5cm]{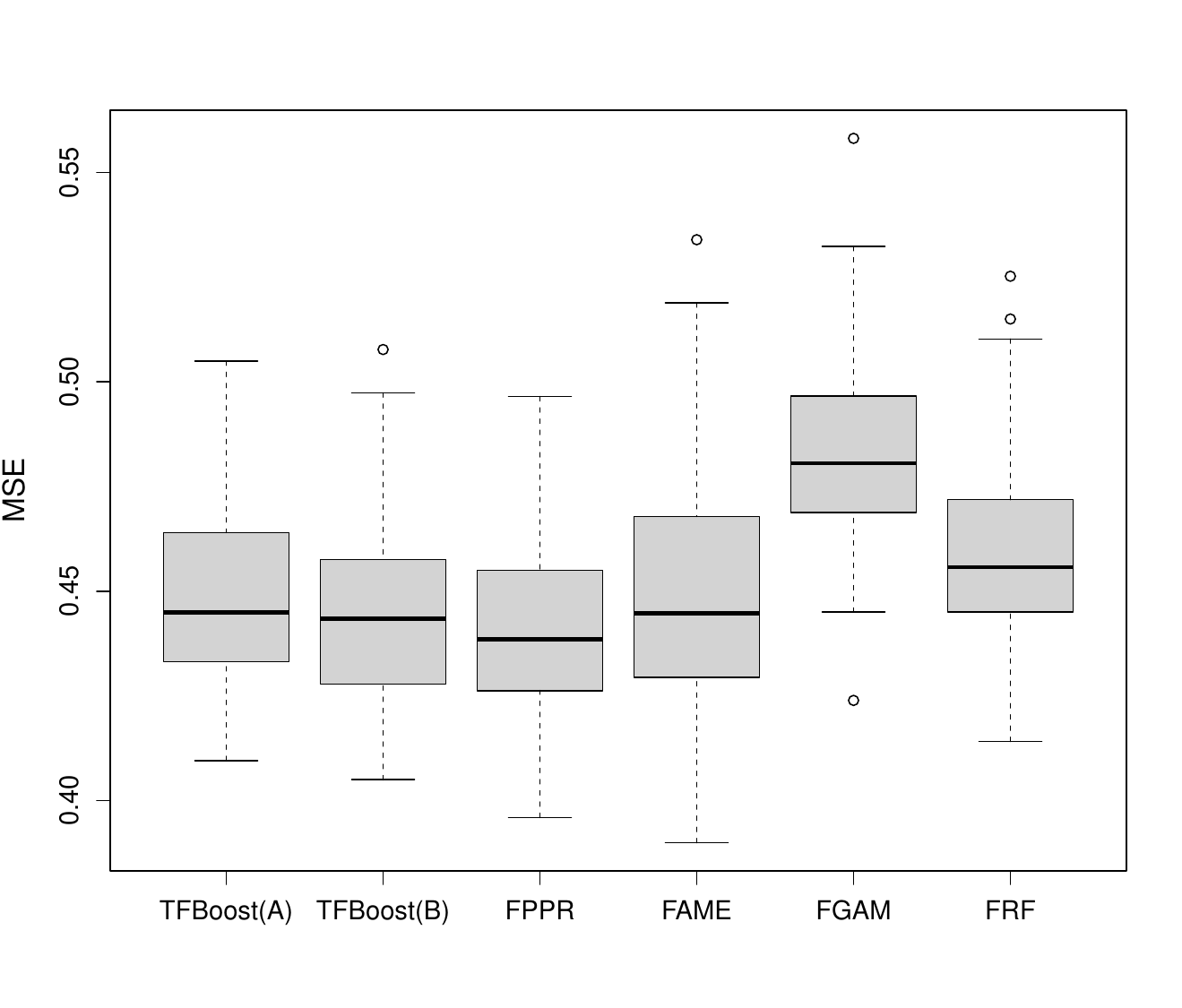}
         \caption{$M_1$}
         \label{}
     \end{subfigure}
     \hfill
     \begin{subfigure}[b]{0.49\textwidth}
         \centering
         \includegraphics[width=1.05\textwidth, height = 4.5cm]{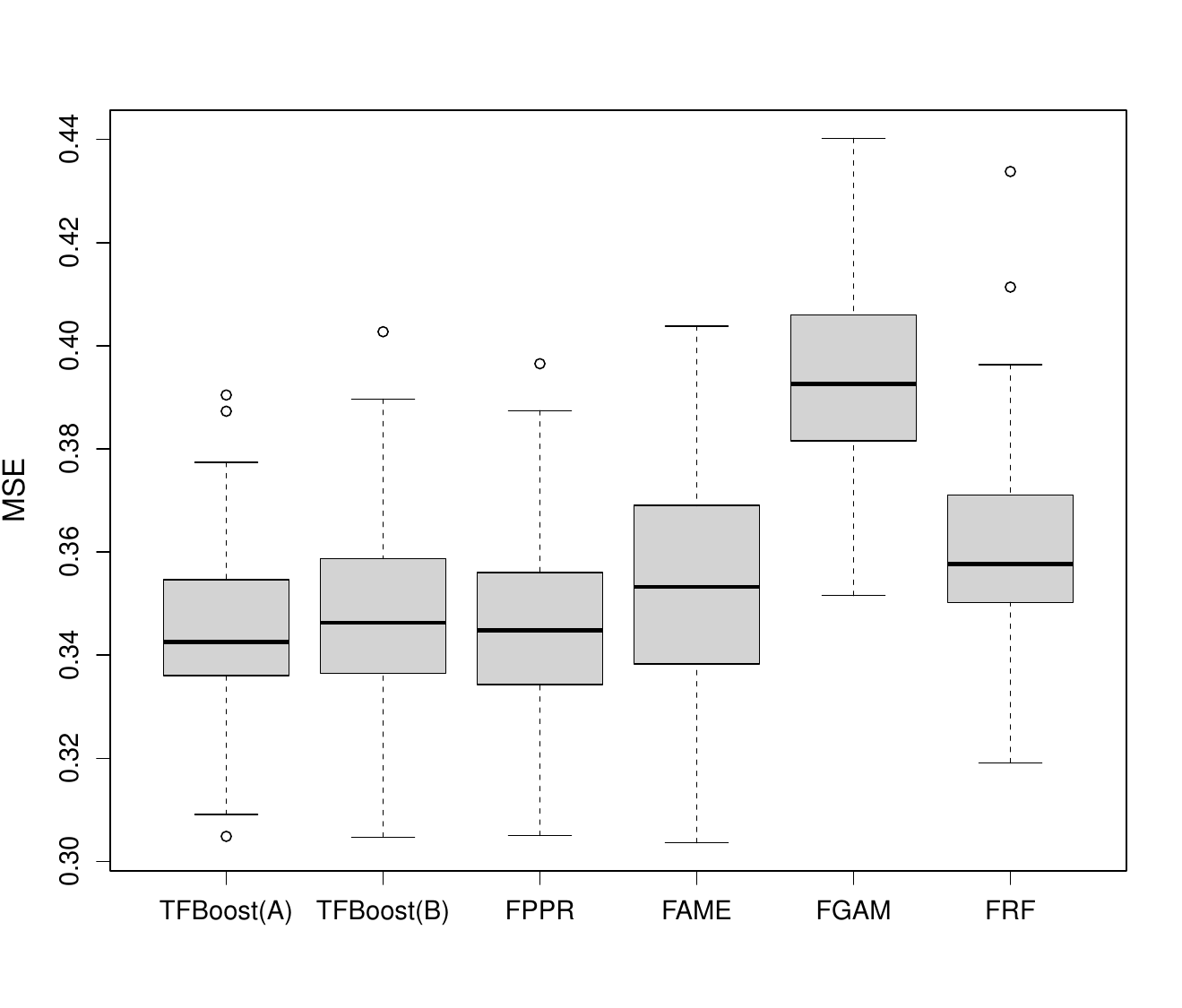}
         \caption{$M_2$}
             \label{}
             
         \label{}
     \end{subfigure}
         \caption{Boxplots of MSEs on test sets from 100 runs of the experiment with data generated from ($r_1$,  $M_1$, $S_1$)  in panel (a) and ($r_1$, $M_2$, $S_1$)  in panel (b).}
        \label{fig:sim1}
\end{figure}

\begin{figure}[htp]
     \centering
     \begin{subfigure}[b]{0.49\textwidth}
         \centering
         \includegraphics[width=1.05\textwidth, height = 4.5cm]{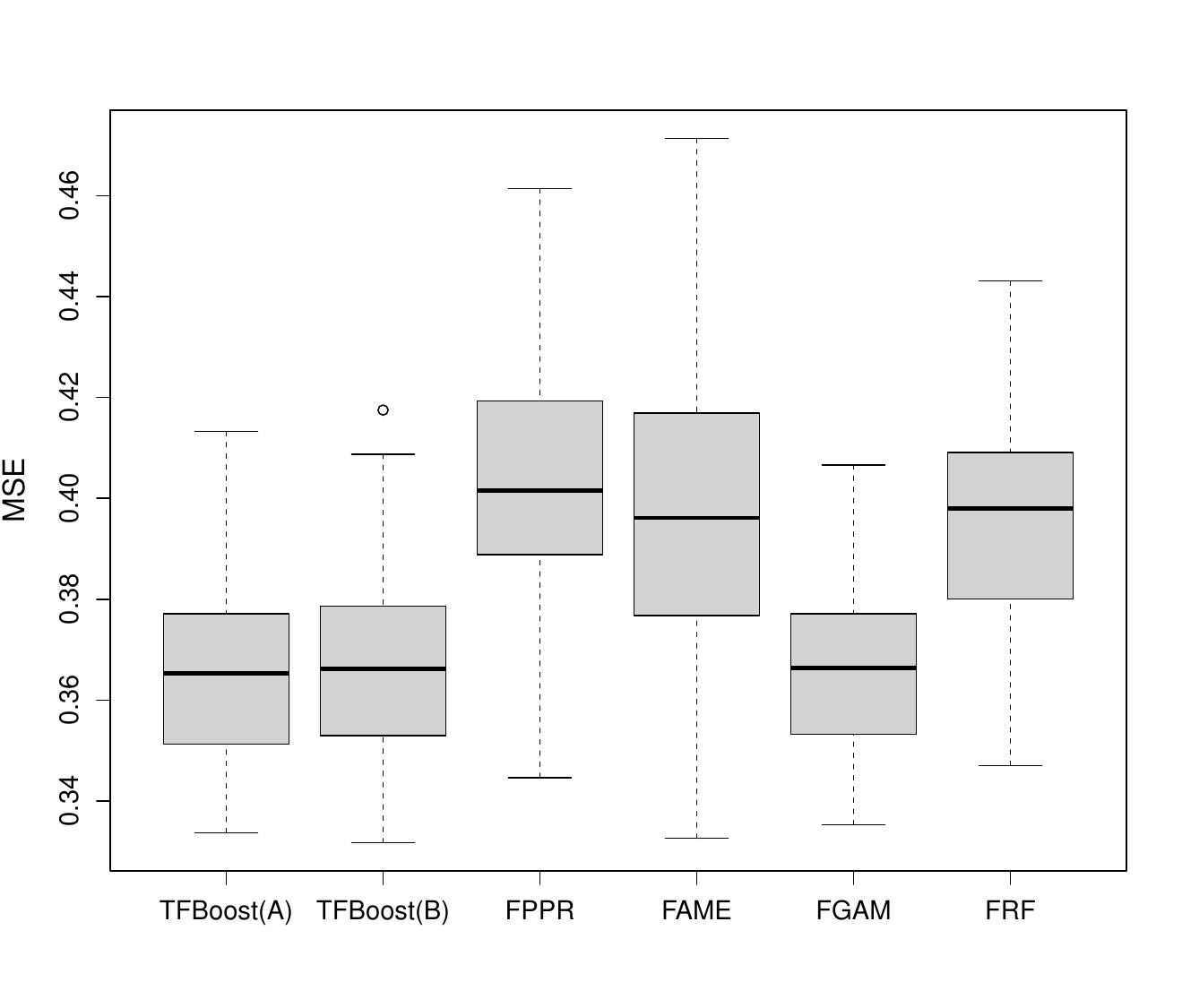}
         \caption{$M_1$}
             \label{}
     \end{subfigure}
     \hfill
     \begin{subfigure}[b]{0.49\textwidth}
         \centering
         \includegraphics[width=1.05\textwidth, height = 4.5cm]{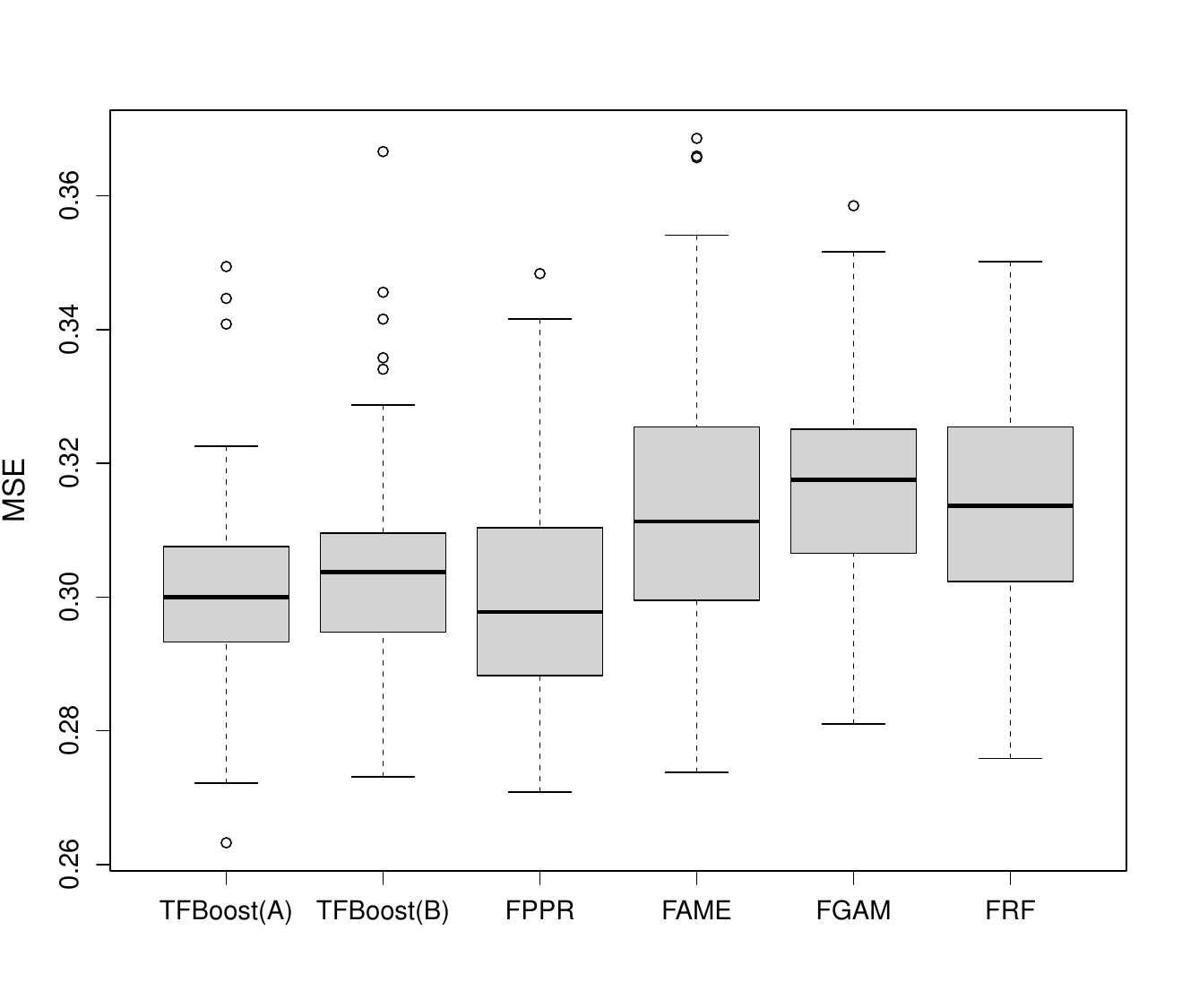} 
         \caption{$M_2$}
          \label{}
     \end{subfigure}
      \caption{Boxplots of MSEs on test sets from 100 runs of the experiment with data generated from ($r_2$,  $M_1$, $S_1$)  in panel (a) and ($r_2$, $M_2$, $S_1$)  in panel (b).}
        \label{fig:sim2}
\end{figure}

\begin{figure}[htp]
     \centering
     \begin{subfigure}[b]{0.49\textwidth}
         \centering
         \includegraphics[width=1.05\textwidth, height = 4.5cm]{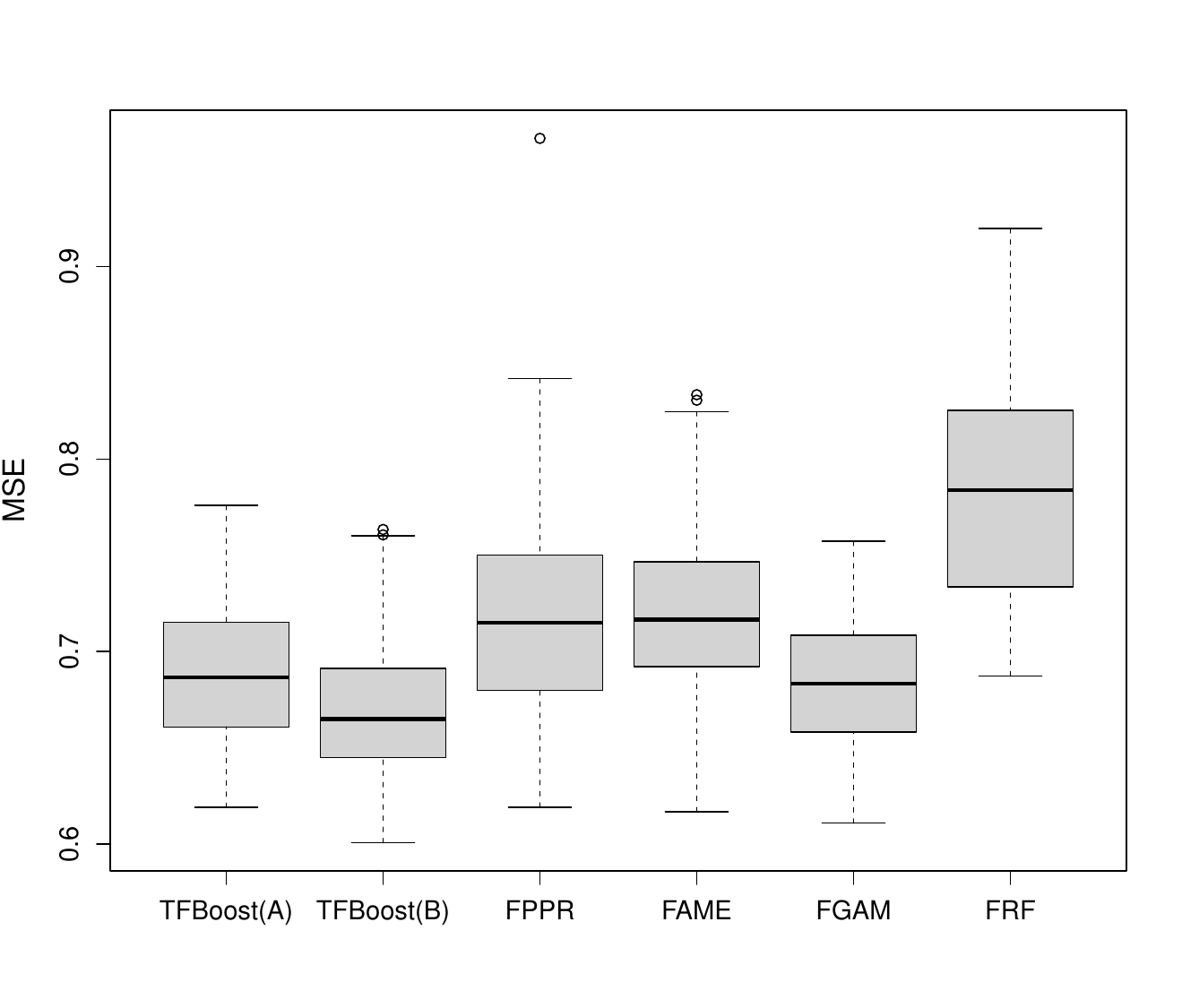}
         \caption{$M_1$}
         \label{}
     \end{subfigure}
     \hfill
     \begin{subfigure}[b]{0.49\textwidth}
         \centering
         \includegraphics[width=1.05\textwidth, height = 4.5cm]{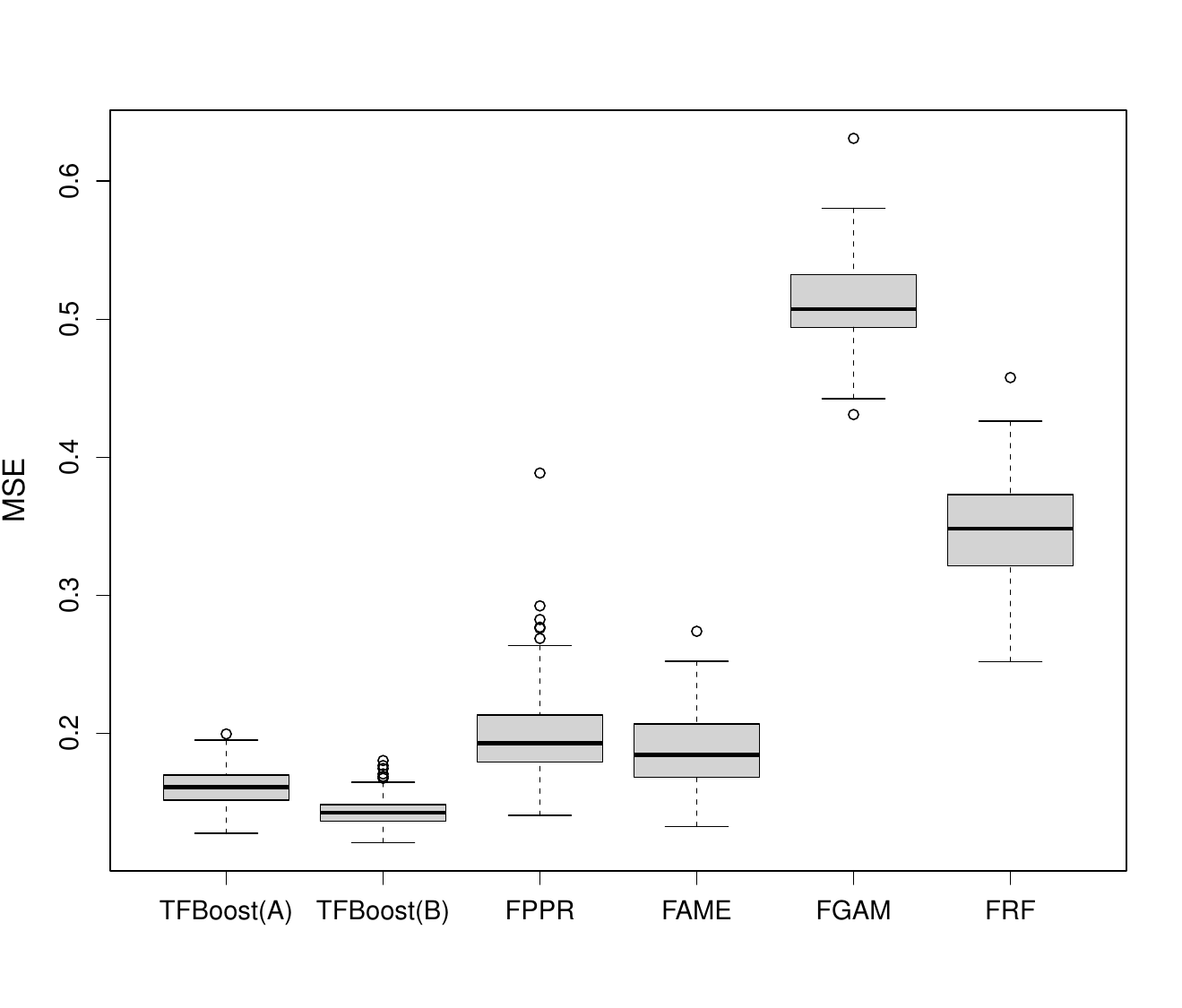}
         \caption{$M_2$}
             \label{}
     \end{subfigure}
      \caption{Boxplots of MSEs on test sets from 100 runs of the experiment with data generated from ($r_3$,  $M_1$, $S_1$)  in panel (a) and ($r_3$, $M_2$, $S_1$)  in panel (b).}
        \label{fig:sim3}
\end{figure}

\begin{figure}[htp]
     \centering
     \begin{subfigure}[b]{0.49\textwidth}
         \centering
         \includegraphics[width=1.05\textwidth, height = 4.5cm]{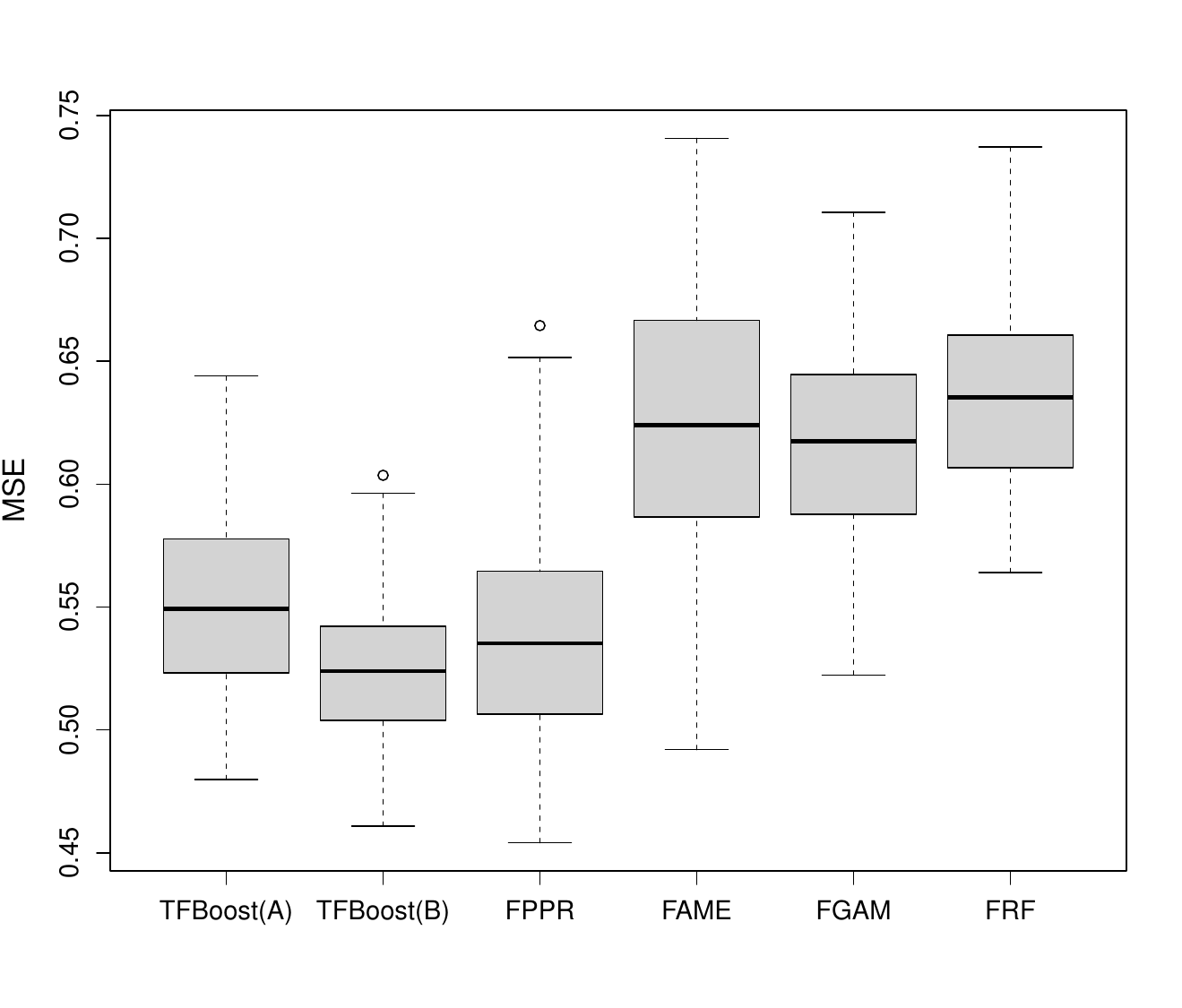}
         \caption{$M_1$}
         \label{}
     \end{subfigure}
     \hfill
     \begin{subfigure}[b]{0.49\textwidth}
         \centering
         \includegraphics[width=1.05\textwidth, height = 4.5cm]{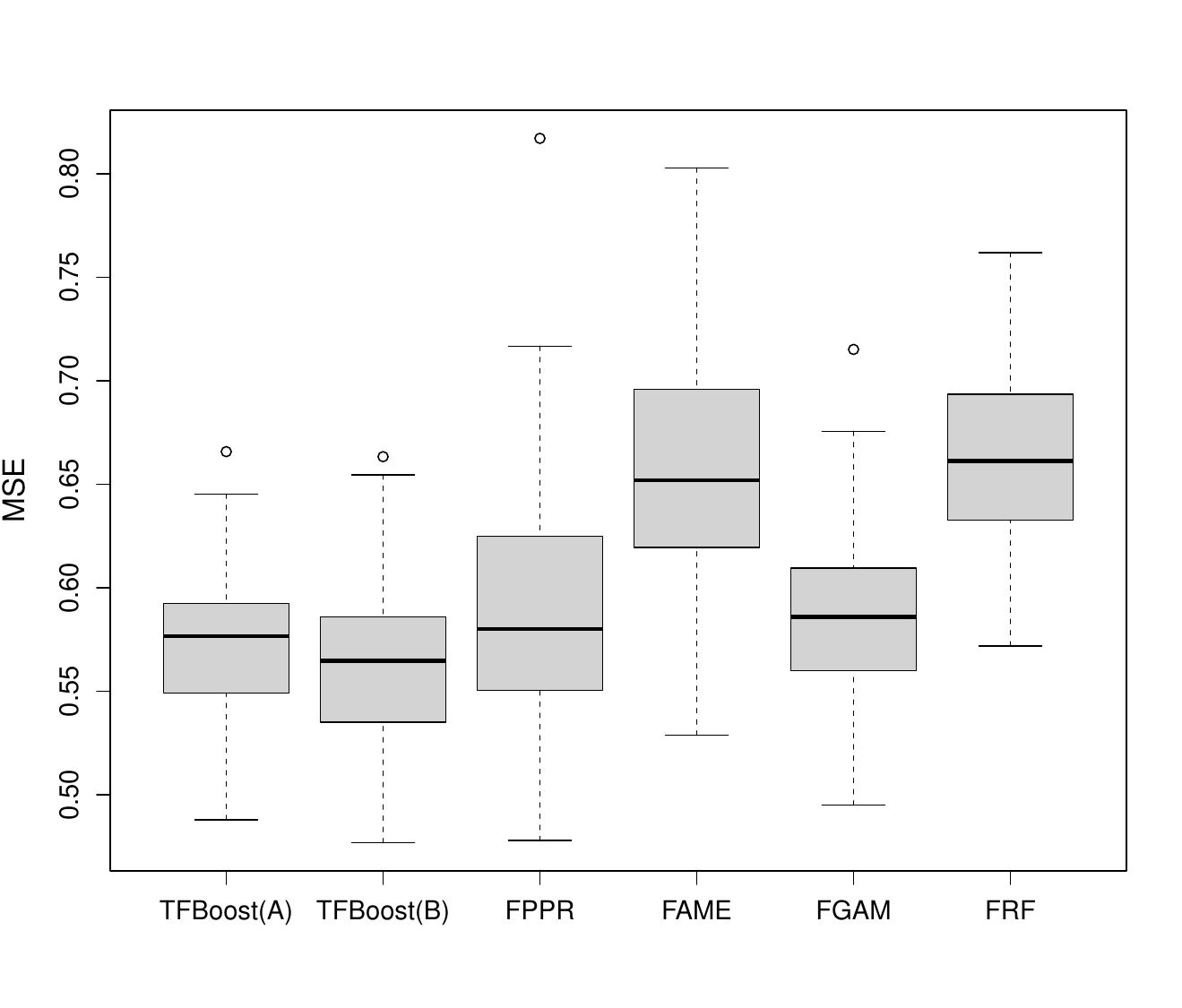}
         \caption{$M_2$}
          \label{}
     \end{subfigure}
      \caption{Boxplots of MSEs on test sets from 100 runs of the experiment with data generated from ($r_4$,  $M_1$, $S_1$)  in panel (a) and ($r_4$, $M_2$, $S_1$)  in panel (b).}
        \label{fig:sim4}
\end{figure}

In each figure, panel (a) corresponds to data with the predictors generated from $M_1$ and panel (b) to those from $M_2$.   Each panel shows the boxplot of MSEs on the test sets
 from 100 independent runs of the experiment. For \texttt{TFBoost(A.K)}, we show the MSEs for the value of $K$ that gave the best results on the test sets, and include the rest in the Appendix.  In the case of a tie we picked the simpler model (one with a smaller $K$).  As a result, we selected 
 
 \begin{itemize}
 \item[- ] $K = 2$ for  all $r_1$ and $r_2$ settings,  and $(r_3, M_2)$,  
 \item[- ]  $K = 3$ for all $r_4$ settings and $(r_3, M_1)$. 
 \end{itemize}
  The label  \texttt{TFBoost(A)} on the x-axis in each figure represents \texttt{TFBoost(A.K)} with  $K$ selected for the corresponding setting.

%We found that the performance of different methods depends on the setting.   No estimator is universally the best in all  settings. 

%In  the linear settings ($r_1, M_1$) and ($r_1, M_2$) shown in \Cref{fig:sim1}, as expected,  \texttt{FLM1}, \texttt{FLM2}, and \texttt{FGAM} outperform the other methods  as their assumptions correspond to the true model.  For  \texttt{FGAM}, we note that the linear model is a particular case of  \ref{eq:fgam} which explains its good performance in these settings.  

%We can see from \Cref{fig:sim1} that the test errors of \texttt{TFBoost(A.1)} are close to those of \texttt{FPPR} and \texttt{FAME}, In terms of the average test MSE, \texttt{TFBoost(A.1)} is about 7\% worse  compared to \texttt{FLM1}, \texttt{FLM2}, and \texttt{FGAM} in  the $M_1$ setting, and about 13\% worse in the $M_2$ setting.   We observed improvements in its test errors when a smaller shrinkage $\gamma$ was used.  For these $r_1$ settings, \texttt{TFBost(B)} prefers $d = 1$ in most runs, and we expect its performance to improve as the number of random directions grows. 

In general,  \texttt{TFBoost(A)}  and  \texttt{TFBoost(B)} show stable and remarkable predictive performance in all these  figures, with either or both achieving the lowest or second lowest average test MSEs across all methods. In contrast,  the performance of \texttt{FPPR}, \texttt{FAME}, and \texttt{FGAM}  heavily depends on the regression function as well as the predictor model. %\texttt{FGAM} modelled the regression function using tensor-product B-splines which multiplies B-splines corresponding to the predictor $X$ and the domain variable $t$. Given B-spline bases  $\{B_j^X(x):j = 1,..., K_x\}$ and $\{B_j^T(t):k = 1,..., K_t\}$, it assumes   $Y = \int_{\mathcal{I}}F\{X_i(t), t \} dt$, where $F(x,t) = \sum_{j=1}^{K_x} \sum_{k=1}^{K_t}\theta_{j,k} B_j^X(x)B_k^T(t)$.  When the true model deviates from this representation, \texttt{FGAM} tend to perform poorly (see \Cref{fig:sim4} for example). 
 For example, in panel (a) of \Cref{fig:sim2} where $X$ was generated from $M_1$, \texttt{FPPR} produces the worst errors compared to the other methods. However, when $X$ followed $M_2$ in panel (b), \texttt{FPPR} becomes one of the best. The opposite holds for \texttt{FGAM}, being among the best in panel (a) and the worst in panel (b). Similarly in  \Cref{fig:sim3}, the performance of \texttt{FGAM} differs greatly in $M_1$ and $M_2$ settings.   In \Cref{fig:sim1,fig:sim4}, the performances of \texttt{FAME} and \texttt{FGAM} are similar in both $M_1$ and $M_2$ settings, but they differ substantially across figures.    \texttt{FGAM}  produces significantly worse errors compared to \texttt{FAME} in \Cref{fig:sim1}, whereas in \Cref{fig:sim4} we observe the opposite. The performance of \texttt{FRF} is at the bottom in all settings, being the worst or the second worst method. We think that this is due to its features constructed using the average of functional values across intervals fail to carry sufficient information to predict the response with high accuracy.

It is worth noting that  
\texttt{TFBoost} methods are close to their alternatives in terms of test errors when the true model matches the one on which the other methods are based. For example, in \Cref{fig:sim1} \texttt{TFBoost(A)} and \texttt{TFBoost(B)} produce similar errors as \texttt{FPPR}, where the regression function $r_2$ matches with the true model of \texttt{FPPR} specified with a single additive component.  In other cases where the alternatives  deviate from the true model, \texttt{TFBoost(A)} and  \texttt{TFBoost(B)} usually show the lowest errors and their performances are stable regardless of the setting. 
This illustrates the flexibility of our method, which  helps to achieve low prediction errors without posing strong assumptions on the target function.

\begin{figure}[htp]
     \centering
     \begin{subfigure}[b]{0.49\textwidth}
         \centering
         \includegraphics[width=1.05\textwidth, page =1]{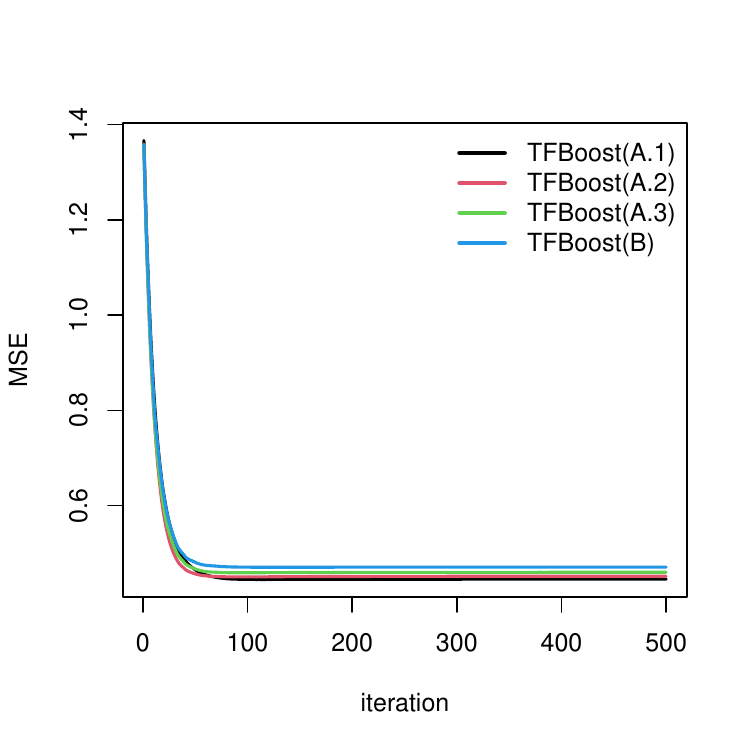}
         \caption{$M_1$}
         \label{}
     \end{subfigure}
     \hfill
     \begin{subfigure}[b]{0.49\textwidth}
         \centering
         \includegraphics[width=1.05\textwidth, page = 2]{figs/testerrors_sim_r1_1.pdf}
         \caption{$M_2$}
          \label{}
     \end{subfigure}
      \caption{Test MSEs averaged from 100 runs of the experiment for \texttt{TFBoost} for data generated  from ($r_1$,  $M_1$, $S_1$)  in panel (a) and ($r_1$, $M_2$, $S_1$)  in panel (b).}
        \label{fig:avg1}
\end{figure}

For each simulated setting, we also analyzed the performance of \texttt{TFBoost} by plotting the test errors averaged from 100 runs of the experiment versus the number of iterations. To make it easier to take the averages for each method,  for each run, we let the test errors for the iterations past the early stopping time ($T_{\text{stop}}$) to have the same value as the one obtained at $T_{\text{stop}}$. \Cref{fig:avg1} shows an example of such figures for  ($r_1$, $M_1$, $S_1$) and ($r_1$, $M_1$, $S_1$)  settings. The summary statistics of $T_{\text{stop}}$ and the selected tree depths $d$ are provided in \Cref{sim:summary}.  We can see that the test errors of \texttt{TFBoost} drop quickly within the first 100 iterations. The same pattern can be observed for the other settings, the results of which  can be found in the Appendix.

\begin{table}[!h]
\centering
\begin{tabular}{rcccc}
 \hline
  & \multicolumn{2}{c}{$M_1$} & \multicolumn{2}{c}{$M_2$}\\  \hline
   & $T_{\text{stop}}$ & $d$ & $T_{\text{stop}}$  & $d$ \\   
  TFBoost(A.1)  & 127.75 (74.22)  & 1.58 (0.77) & 149.05 (130.43) & 1.77 (0.98) \\ 
  TFBoost(A.2) &  121.02 (82.62) & 1.55 (0.78) & 120.74 (76.94) & 1.67 (0.96) \\ 
  TFBoost(A.3)  & 113.52 (54.25) & 1.40 (0.65) & 121.83 (94.96) & 1.64 (0.82) \\ 
  TFBoost(B)  & 96.05 (38.05)  & 1.32 (0.66) & 107.63 (50.49) & 1.32 (0.60) \\ 
   \hline
\end{tabular}
\caption{Summary statistics of the early stopping times ($T_{\text{stop}}$) and tree depths ($d$) selected by TFBoost for data generated from ($r_1$, $M_1$, $S_1$) and ($r_1$, $M_2$, $S_1$); displayed in the form of mean (sd).} 
\label{sim:summary}
\end{table}

So far we have focused on the performance of \texttt{TFBoost} in nonlinear settings since the base learners are nonlinear. 
In the Appendix, we provide also results for an experiment fitting \texttt{TFBoost} methods with data generated from a linear model and comparing them with alternatives considered previously.  As expected,  the linear estimators \texttt{FLM1}, \texttt{FLM2} outperform the others.  For \texttt{FGAM}, the linear regression function matches with its model assumptions, which explains its good performance in linear settings.  We note that the test errors of \texttt{TFBoost(A)} are only slightly worse than those produced by linear estimators:  about 5\% worse in  the $M_1$ setting, and about 7\% worse in the $M_2$ setting.  This implies that even though \texttt{TFBoost} is an estimator for possibly nonlinear regression functions, we do not lose much applying it to data generated from  linear models.

\section{German electricity data}
\label{sec:real}
To illustrate the use of \texttt{TFBoost}, we analyzed a data set consisting of electricity spot prices traded at the European Energy Exchange (EEX) in Leipzig and electricity demand reported by the European Network of Transmission System Operators for Electricity from January 1st 2006 to  September 30th 2008. 
 On each day,  electricity spot prices and demands were recorded hourly and represented as vectors of dimension 24.    Our objective is to predict the daily average of electricity demand using  hourly electricity spot prices.    The whole dataset is available from the on-line supplementary materials of \cite{liebl2013modeling}.   To avoid days with known atypical price or demand values affecting our analysis, we removed weekends and holidays and analyzed data collected on the remaining 638 days.  For each of these days, we view the hourly electricity spot prices  as discretized evaluations of a smooth price function.  

We considered methods of  \texttt{TFBoost(A,K)} with $K$ = 1, 2, and 3, \texttt{TFBoost(B)}, as well as their alternatives \texttt{FLM1}, \texttt{FLM2}, \texttt{FAM}, \texttt{FPPR}, \texttt{FAME},  \texttt{FGAM}, \texttt{RFgroove}, and \texttt{FRF}.  To adjust for  potential seasonality effects, we created a ``day of the year'' variable defined as the number of days from January 1st of the year in which the data were collected. In order to compare the results before and after adjusting for seasonality, our analysis consisted of two parts: (1)  predicting daily electricity demand with 
 hourly electricity prices only,  and (2) with both hourly electricity  prices and the ``day of the year'' variable.  
 
 In part (1),  all methods under consideration were specified in the same way as in \Cref{sec:imp}.   In part (2), we added ``day of the year"  as an additional predictor to all methods  and adjusted their algorithms accordingly. For \texttt{TFBoost} methods, at every boosting iteration, we added ``day of the year" as an additional variable to the functional multi-index tree.  For the competitors,  we included the scalar predictor in the model following the same approach as we did for the functional predictor. We added ``day of the year"  as a linear term in \texttt{FLM1} and \texttt{FLM2}, and as a nonparametric term in \texttt{FAM}, \texttt{FPPR}, \texttt{FAME}, \texttt{FGAM}, \texttt{RFgroove}, and \texttt{FRF}.  For each of \texttt{FAM} and \texttt{FPPR}, we first calculated the estimator in the same way as in part (1) and obtained the residuals.  Then we added to the part (1) estimator  a Nadaraya-Watson estimator with Gaussian kernel fitted to the residuals using ``day of the year"  as the predictor.   For \texttt{FAME} and \texttt{FGAM} we specified a smooth term on ``day of the year"  represented as penalized cubic splines \citep{wood2017generalized}. For \texttt{RFgroove} and \texttt{FRF}, we added ``day of the year"  as an additional predictor to each regression tree used to construct random forests.   Other than adding this additional predictor, the other parameters and fitting procedures were kept the same as in part (1).

We randomly partitioned the data into a training set (60\%), a validation set (20\%), and a test set (20\%).  As in \Cref{Sec:sim},  we fitted each model using the training and validation sets and recorded MSE on the test set.     The validation set was used to choose the regularization parameter of \texttt{FLM1}, the number of additive components of \texttt{FAM} and \texttt{FPPR}, the rate parameters to generate the intervals for \texttt{FRF}, and the maximum tree depth and  early stopping time of \texttt{TFBoost(A,1)},\texttt{TFBoost(A,2)}, \texttt{TFBoost(A,3)},  and \texttt{TFBoost(B)}.

\begin{figure}[H]
     \centering
 \includegraphics[width=\textwidth, height = 6.5cm, page=1]{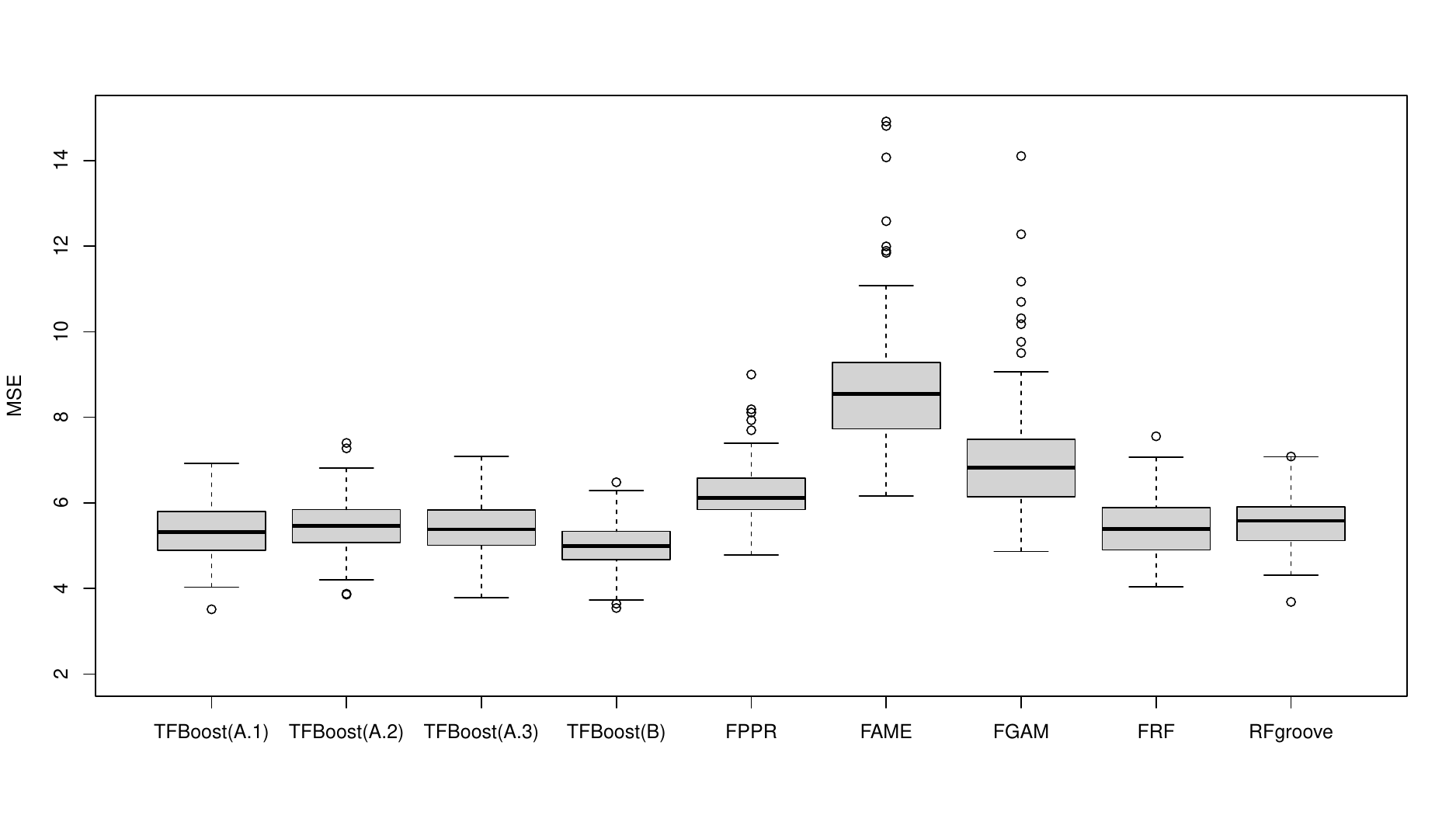}
\caption{Boxplot of test MSEs obtained from 100 random splits of the data from part~(1) of the experiment. The unit of y-axis is $10^{-6}$.}
\label{fig:r1}
\end{figure}

\begin{figure}[H]
     \centering
 \includegraphics[width=\textwidth, height = 6.5cm, page=2]{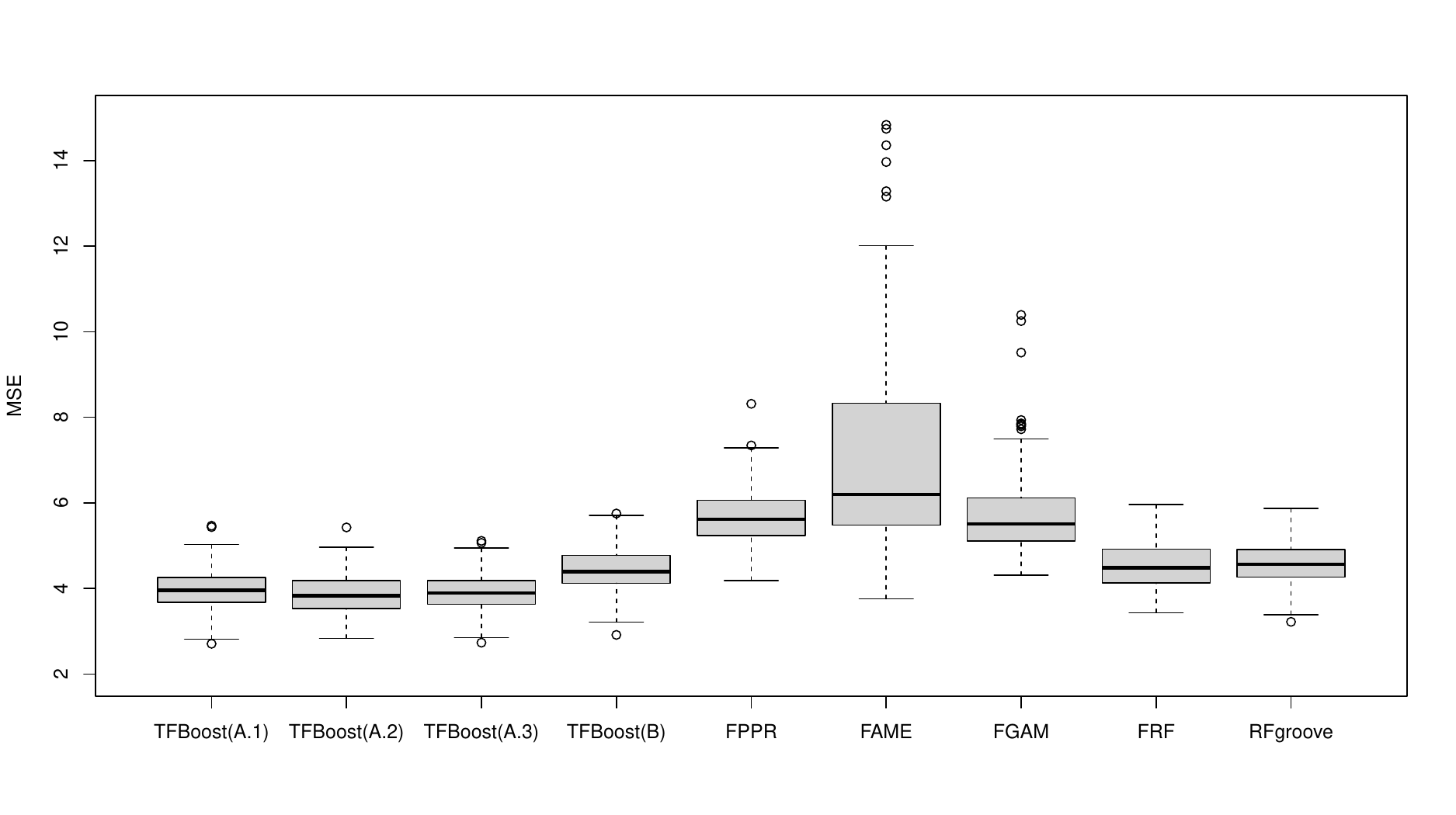}
\caption{Boxplot of test MSEs obtained from 100 random splits of the data from part~(2) of the experiment. The unit of y-axis is $10^{-6}$.}
\label{fig:r2}
\end{figure}

To reduce the variability introduced by partitioning the data, we repeated the experiment with 100 random data splits. \Cref{fig:r1,fig:r2}  show test MSEs obtained from 100 random data splits in part (1) and (2) of the experiment respectively. The summary statistics of these test MSEs for all methods are provided in the Appendix.   To better reveal the differences between the estimators under consideration, in each figure the y-axis is truncated and represents test MSE $\times 10^{-6}$.  We also excluded \texttt{FLM1}, \texttt{FLM2}, and \texttt{FAM} from the figures since they performed very poorly in part(1) and part(2).

 In both figures, \texttt{TFBoost(A.1)}, \texttt{TFBoost(A.2)},  \texttt{TFBoost(A.3)}, and  \texttt{TFBoost(B)} show superior performance compared to their alternatives. They achieve the lowest errors with the smallest standard errors.    In \Cref{fig:r1},  we  observe that \texttt{TFboost(B)} produces the smallest test errors, substantially better than \texttt{FLM1},\texttt{FLM2}, \texttt{FAM},  \texttt{FGAM}, and \texttt{FPPR}, and slightly better than \texttt{FRF} and \texttt{RFgroove}. 
  
 Comparing \Cref{fig:r2} with \Cref{fig:r1},  we observe improvements in test errors for all methods except for \texttt{FLM1}. This implies the importance of adjusting for seasonality when predicting the electricity demand, which is likely due to higher electricity usage in the summer for cooling and in the winter for heating. 
  In \Cref{fig:r2},    \texttt{TFBoost(A.1)}, \texttt{TFBoost(A.2)}, \texttt{TFBoost(A.3)}  produce very similar results, and \texttt{TFBoost(A.2)} outperforms the other two by a small margin in terms of the average test MSE. 
 
%Its performance improves as tree grows deeper, suggesting the importance of  modelling complex index interactions for improving prediction accuracy.  \texttt{TFBoost(A.1)}, \texttt{TFBoost(A.2)}, and \texttt{TFBoost(A.3)} perform better with shallow trees ($d = 1$ or $2$) and show overfitting as  trees grow deeper.  In , we plot the test MSEs of \texttt{TFBoost(A.1)}, \texttt{TFBoost(A.2)}, \texttt{TFBoost(A.3)}, and \texttt{TFBoost(B)} at the tree depth that resulted in better performances ($d = 1,1,2$ and 4 respectively) as well as the test MSEs of their competitors. 

% We observe that \texttt{TFBoost} methods produce the lowest test errors with small variances.  The performance of \texttt{FPPR} is the closest to \texttt{TFBoost(B)}, however, its computation takes much longer considering its calculation of the Gram matrix and selection of the bandwidth when constructing kernel estimators.  The computation budget can seriously exacerbate as the sample size grows, which contrasts with \texttt{TFBoost(B)} where the computation cost to fit a tree is proportional to the sample size. 

\section{Conclusion and future work} \label{Sec:con and future work}

In this paper we have proposed a novel boosting method for regression with a functional explanatory variable. Our approach uses functional multi-index trees as  base learners, which offer flexibility and are relatively easy to fit. Compared with widely studied single index estimators, our multi-index estimator enables modelling possible interactions between indices, allowing us to better approximate complex regression functions.  Through extensive numerical experiments, we demonstrate that our method consistently produces one of the lowest prediction errors across different settings, while available alternatives can be  seriously affected by model misspecification. In addition, \texttt{TFBoost} with Type B learners exhibits significant computational advantages over kernel-based methods, being able to provide accurate predictions at a much lower cost.

The methodology of \texttt{TFBoost} suggests a number of interesting areas for future work. First, we formulated the regression problem with a single functional predictor, but in principle, it could also be extended to multiple predictors, functional or scalar.  This can be achieved by including indices calculated from multiple predictors, either allowing each index to be associated with all predictors or one of the predictors.  If there are many predictors, this approach may be  computationally  difficult and can result in unstable estimators when the sample size is small.  In this situation, we need to assume sparsity in the functional space, which means that some but not all predictors are related to the response. Group penalties with projections on each predictor being a group can be considered to enable variable selection.  Second, in our experiments we only studied \texttt{TFBoost} applied with the squared loss. In cases where data contain outliers, it will be useful to consider fitting  our method 
 with different loss functions, for example, Huber's or Tukey's loss.  Finally, in this paper we focused on functional predictors observed at many time points without measurement errors. In practice, one may encounter situations where only very few and possibly noisy measurements of the predictor function are available.  In this case, smoothing each predictor curve provides unreliable approximations and makes them poor inputs to \texttt{TFBoost}. It may be more advisable to borrow strength across curves, for example,  fitting \texttt{TFBoost} to curves reconstructed using FPCA  (\cite{yao2005functional} and \cite{boente2021robust}). 
 
\section{Acknowledgements}
The authors would like to thank Professors James and Ferraty for sharing the code used in their papers (\cite{james2005functional} and \cite{ferraty2013functional}). In addition, 
we would like to thank two anonymous referees and an Associate Editor for their
constructive comments on an earlier version of this work that resulted in a notably
improved paper. 

\section{Statements and Declarations}
This research was supported by the Natural Sciences and Engineering Research Council of Canada [Discovery Grant RGPIN-2016-04288]. The authors have no competing interests to declare that are relevant to the content of this article.

\newpage 
\bibliographystyle{apalike} 
\bibliography{mybibfile.bib}

\newpage
\bigskip
\begin{center}
{\large\bf APPENDICES}
\end{center}

Proof of \Cref{thm:1} in \Cref{subsec:typeA}. 
\begin{proof} \label{prop:iden}
It is clear that if $\{\beta_1,..., \beta_K \} = \{(-1)^{l_1}\eta_1,..., (-1)^{l_K}\eta_K \}$ for some $l_1,...,l_K \in \{0,1\}$, then  $g = \tilde{g}$.  Therefore, it suffices to show that  $\{\beta_1,..., \beta_K \} = \{(-1)^{l_1}\eta_1,..., (-1)^{l_K}\eta_K \}$ for some $l_1,...,l_K \in \{0,1\}$.  We prove that if there do not exist $l_1,...,l_K$ for the two sets to be equal, there exists a  set of indices for which  \eqref{eq:proof} is a constant function and thus contradicts \Cref{condition:iden2}. 

For simplicity, we let  $\tilde{\eta}_j = (-1)^{l_j} \eta_j$. If for any $l_1,...,l_K$, $\{\beta_1,..., \beta_K \} \neq \{\tilde{\eta}_1,..., \tilde{\eta}_K \}$, we match two sets so that the same vectors $\beta_j$ and $\tilde{\eta}_j$ align with each other. We let $S = \{\beta_1,...,\beta_K \} \cap \{\tilde{\eta}_1,...,\tilde{\eta}_K \}$, $\beta_j = \tilde{\eta}_j$, for $j = 1,..., |S|$ and $\beta_j \notin  \{\tilde{\eta}_1,...,\tilde{\eta}_K \}$, for $j = |S|+1,..., K$, and $|S| < K$.   By \Cref{condition:iden2}, there exist a $x_0$ for $J = {|S|+1,...,K}$,  \eqref{eq:proof} is not a constant function.

  By \eqref{eq:thm}, \Cref{condition:iden1} and \Cref{condition:iden2}, for any $t_1,..., t_K \in (-\delta, \delta)$  \begin{align*}
h( \langle  x_0 ,\beta_1\rangle + t_1, ..., \langle x_0, \beta_K \rangle + t_K)	 &= h(\langle   x_0 + t_1\beta_1, \beta_1  \rangle, ..., \langle  x_0 + t_K \beta_K\rangle, \beta_K ) \\
&=\tilde{h} (\langle  x_0 + t_1 \beta_1, \eta_1  \rangle, ..., \langle x_0 + t_K \beta_K, \eta_K \rangle ) \\
&= \tilde{h} ( \langle  x_0, \eta_1  \rangle + t_1 \langle  \beta_1,  \eta_1\rangle, ..., \langle x_0, \eta_K \rangle  + t_K \langle \beta_K,\eta_K\rangle )
\end{align*}
and similarly 
\begin{align*}
\tilde{h}( \langle x_0, \eta_1 \rangle + t_1, ..., \langle x_0, \eta_K \rangle + t_K)	 &= \tilde{h}(\langle  x_0 + t_1 \eta_1 , \eta_1 \rangle, ..., \langle x_0 + t_K  \eta_K, \eta_K\rangle ) \\
&=h (\langle x_0 + t_1 \eta_1, \beta_1  \rangle, ..., \langle x_0 + t_K \eta_K, \beta_K \rangle ) \\
&= h ( \langle  x_0,  \beta_1 \rangle + t_1 \langle \beta_1, \eta_1 \rangle, ..., \langle x_0, \beta_K \rangle  + t_K \langle \beta_K, \eta_K\rangle )
\end{align*}

By Cauchy-Schwarz inequality and \Cref{condition:iden1},  $(\langle \beta_j, \eta_j \rangle)^2 = 1$ for $j = 1, ..., |S|$ and $(\langle \beta_j, \eta_j \rangle)^2 < 1$ for $j = |S+1|, ..., K$.  For any $t_1,..., t_K \in (-\delta,\delta)$, 
\begin{align} \label{myeq}
h( \langle x_0,  \beta_1 \rangle + t_1, ..., \langle x_0 , \beta_K \rangle + t_K)  \nonumber  
	&= \tilde{h} ( \langle x_0,\eta_1  \rangle + t_1 \langle  \beta_1, \eta_1 \rangle, ..., \langle x_0, \eta_K \rangle  + t_K \langle \beta_K, \eta_K\rangle ) \nonumber \\
	&=  h ( \langle x_0,  \beta_1 \rangle + t_1 \langle \beta_1, \eta_1 \rangle^2, ..., \langle  x_0, \beta_K\rangle  + t_K \langle \beta_K,  \eta_K\rangle^2 ) \nonumber \\
	& \vdots  \nonumber \\
	&= h ( \langle x_0, \beta_1  \rangle + t_1 \langle \beta_1, \eta_1 \rangle^{2n}, ..., \langle  x_0, \beta_K\rangle  + t_K \langle \beta_K,  \eta_K \rangle^{2n} )\nonumber  \\
	& \vdots \nonumber \\
	&= h ( \langle  x_0, \beta_1  \rangle + t_1I_1, ..., \langle  x_0, \beta_K\rangle  + t_KI_K) 
\end{align}
where $I_j = 1$ for $j = 1,...,|S|$ and $I_j = 0$ for $j = |S+1|,..., |K|$.

Let $x = x_0 + t e$ for any unit function  $e \in L^2(\mathcal{I})$, $\lVert e \rVert = 1$ and $t \in (-\delta, \delta)$.  Then $x$ fills the space of $B(x_0, \delta)$. For $j = 1,...,K$, we define  
\begin{align*}
    L_j(x) &= (1-I_j)x + I_j x_0 \\
    &= (1-I_j) (x_0 + te) + I_j x_0 \\
    &= x_0 + (1 - I_j) te
\end{align*}
\begin{align*}
		h( \langle  L_1(x),\beta_1 \rangle,..., \langle  L_K(x), \beta_K \rangle) &= 	h( \langle  x_0 + (1 - I_1) te, \beta_1 \rangle,...,\langle x_0 + (1 - I_K) te, \beta_K  \rangle) \\
		& = h( \langle  x_0, \beta_1 \rangle +  \langle  (1 - I_1) te, \beta_1 \rangle,..., \langle x_0, \beta_K \rangle +  \\   & \langle  (1 - I_K) te, \beta_K \rangle) , \text{by} \ \eqref{myeq}  \\
&= h( \langle  x_0, \beta_1 \rangle +  \langle   I_1 (1 - I_1) te, \beta_1 \rangle,..., \langle  x_0, \beta_K \rangle + \\ &  \langle I_K(1 - I_K) te, \beta_K \rangle) \\
&=h( \langle x_0, \beta_1 \rangle,..., \langle x_0, \beta_K \rangle ), 
\end{align*}
which is a constant function of $x$ and that contradicts \Cref{condition:iden2}.
\end{proof}

\section*{Appendix B}

The summary statistics of test MSEs from 100 independent runs of the simulation are provided in \Cref{er:1,er:2,er:3,er:4}, with bold font indicating the lowest two average test errors in each setting. Summary statistics of the tree depths selected by \texttt{TFBoost} are provided in  \Cref{dd:1,dd:2,dd:3,dd:4}, and for the early stopping times for \texttt{TFBoost} are provided in  \Cref{e:1,e:2,e:3,e:4}.

% latex table generated in R 4.2.2 by xtable 1.8-4 package
% Fri Jan 13 00:27:47 2023
\begin{table}[!h]
\centering
\begin{tabular}{rcccc}
  \hline
  & \multicolumn{2}{c}{SNR = 20} & \multicolumn{2}{c}{SNR = 5}\\ \hline
 & $M_1$ & $M_2$ & $M_1$ & $M_2$ \\ 
  TFBoost(A.1) & 0.118 (0.006) & \textbf{0.090} (0.004) & 0.449 (0.021) & \textbf{0.347} (0.016) \\ 
  TFBoost(A.2) & 0.118 (0.006) & \textbf{0.090} (0.004) & 0.449 (0.022) & \textbf{0.345} (0.016) \\ 
  TFBoost(A.3) & 0.120 (0.005) & \textbf{0.090} (0.004) & 0.450 (0.020) & \textbf{0.347} (0.016) \\ 
  TFBoost(B) & \textbf{0.116} (0.005) & 0.091 (0.004) & \textbf{0.445} (0.022) & \textbf{0.347} (0.016) \\ 
  FLM1 & 0.218 (0.009) & 0.200 (0.008) & 0.534 (0.022) & 0.445 (0.019) \\ 
  FLM2 & 0.219 (0.008) & 0.200 (0.008) & 0.534 (0.021) & 0.446 (0.019) \\ 
  FAM & 0.308 (0.055) & 0.190 (0.010) & 0.624 (0.061) & 0.438 (0.021) \\ 
  FPPR & \textbf{0.113} (0.006) & \textbf{0.089} (0.005) & \textbf{0.440} (0.021) & \textbf{0.347} (0.018) \\ 
  FAME & 0.118 (0.008) & 0.094 (0.007) & 0.448 (0.026) & 0.353 (0.021) \\ 
  FGAM & 0.151 (0.007) & 0.134 (0.006) & 0.483 (0.022) & 0.393 (0.018) \\ 
  FRF & 0.138 (0.008) & 0.113 (0.006) & 0.459 (0.022) & 0.360 (0.019) \\ 
  RFGroove & 0.179 (0.011) & 0.165 (0.010) & 0.521 (0.023) & 0.423 (0.021) \\ 
   \hline
\end{tabular}
\caption{Summary statistics of test errors for data generated from $r_1$; displayed in the form of mean (sd).} 
\label{er:1}
\end{table}

% latex table generated in R 4.2.2 by xtable 1.8-4 package
% Fri Jan 13 00:27:48 2023
\begin{table}[!h]
\centering
\begin{tabular}{rcccc}
  \hline
  & \multicolumn{2}{c}{SNR = 20} & \multicolumn{2}{c}{SNR = 5}\\ \hline
 & $M_1$ & $M_2$ & $M_1$ & $M_2$ \\ 
  TFBoost(A.1) & 1.59 (0.64) & 1.98 (0.88) & 1.58 (0.77) & 1.77 (0.98) \\ 
  TFBoost(A.2) & 1.44 (0.61) & 1.71 (0.84) & 1.55 (0.78) & 1.67 (0.96) \\ 
  TFBoost(A.3) & 1.54 (0.63) & 1.74 (0.79) & 1.40 (0.65) & 1.64 (0.82) \\ 
  TFBoost(B) & 1.51 (0.73) & 1.52 (0.66) & 1.32 (0.66) & 1.32 (0.60) \\ 
   \hline
\end{tabular}
\caption{Summary statistics of the tree depths selected by TFBoost for data generated from $r_1$; displayed in the form of mean (sd).} 
\label{dd:1}
\end{table}
% latex table generated in R 4.2.2 by xtable 1.8-4 package
% Fri Jan 13 00:27:49 2023
\begin{table}[!h]
\centering
\begin{tabular}{rcccc}
  \hline
  & \multicolumn{2}{c}{SNR = 20} & \multicolumn{2}{c}{SNR = 5}\\ \hline
 & $M_1$ & $M_2$ & $M_1$ & $M_2$ \\ 
  TFBoost(A.1) & 184.36 (148.77) & 179.12 (140.80) & 127.75 (74.22) & 149.05 (130.43) \\ 
  TFBoost(A.2) & 159.15 (104.41) & 160.64 (124.48) & 121.02 (82.62) & 120.74 (76.94) \\ 
  TFBoost(A.3) & 151.91 (68.41) & 160.39 (122.58) & 113.52 (54.25) & 121.83 (94.96) \\ 
  TFBoost(B) & 120.24 (50.50) & 140.08 (104.60) & 96.05 (38.05) & 107.63 (50.49) \\ 
   \hline
\end{tabular}
\caption{Summary statistics of the early stoping times $T_{\text{stop}}$  selected by TFBoost methods for data generated from $r_1$; displayed in the form of mean (sd).} 
\label{e:1}
\end{table}
% latex table generated in R 4.2.2 by xtable 1.8-4 package
% Fri Jan 13 00:27:50 2023
\begin{table}[!h]
\centering
\begin{tabular}{rcccc}
  \hline
  & \multicolumn{2}{c}{SNR = 20} & \multicolumn{2}{c}{SNR = 5}\\ \hline
 & $M_1$ & $M_2$ & $M_1$ & $M_2$ \\ 
  TFBoost(A.1) & 0.143 (0.009) & \textbf{0.091} (0.005) & 0.371 (0.018) & 0.302 (0.014) \\ 
  TFBoost(A.2) & 0.133 (0.008) & \textbf{0.089} (0.005) & \textbf{0.365} (0.018) & \textbf{0.301} (0.014) \\ 
  TFBoost(A.3) & \textbf{0.130} (0.009) & \textbf{0.089} (0.005) & \textbf{0.365} (0.019) & \textbf{0.301} (0.014) \\ 
  TFBoost(B) & \textbf{0.124} (0.008) & \textbf{0.091} (0.005) & \textbf{0.366} (0.019) & 0.304 (0.015) \\ 
  FLM1 & 0.610 (0.032) & 0.177 (0.011) & 0.815 (0.040) & 0.376 (0.018) \\ 
  FLM2 & 0.611 (0.032) & 0.176 (0.011) & 0.816 (0.040) & 0.376 (0.017) \\ 
  FAM & 0.352 (0.076) & 0.137 (0.012) & 0.560 (0.081) & 0.337 (0.017) \\ 
  FPPR & 0.155 (0.014) & 0.092 (0.006) & 0.402 (0.024) & \textbf{0.300} (0.017) \\ 
  FAME & 0.148 (0.016) & 0.099 (0.010) & 0.397 (0.029) & 0.313 (0.020) \\ 
  FGAM & 0.150 (0.009) & 0.113 (0.008) & \textbf{0.366} (0.017) & 0.316 (0.014) \\ 
  FRF & 0.192 (0.011) & 0.118 (0.008) & 0.394 (0.020) & 0.313 (0.015) \\ 
  RFGroove & 0.216 (0.012) & 0.178 (0.015) & 0.433 (0.021) & 0.382 (0.021) \\ 
   \hline
\end{tabular}
\caption{Summary statistics of test errors for data generated from $r_2$; displayed in the form of mean (sd).} 
\label{er:2}
\end{table}
% latex table generated in R 4.2.2 by xtable 1.8-4 package
% Fri Jan 13 00:27:51 2023
\begin{table}[!h]
\centering
\begin{tabular}{rcccc}
  \hline
  & \multicolumn{2}{c}{SNR = 20} & \multicolumn{2}{c}{SNR = 5}\\ \hline
 & $M_1$ & $M_2$ & $M_1$ & $M_2$ \\ 
  TFBoost(A.1) & 1.75 (0.72) & 2.17 (1.02) & 1.79 (0.83) & 2.20 (1.06) \\ 
  TFBoost(A.2) & 2.72 (0.81) & 2.85 (0.90) & 2.46 (1.09) & 2.35 (1.05) \\ 
  TFBoost(A.3) & 3.07 (0.82) & 2.95 (0.94) & 2.53 (1.05) & 2.35 (0.91) \\ 
  TFBoost(B) & 3.34 (0.68) & 2.66 (1.00) & 2.72 (1.05) & 1.99 (0.92) \\ 
   \hline
\end{tabular}
\caption{Summary statistics of the tree depths selected by TFBoost  for data generated from $r_2$; displayed in the form of mean (sd).} 
\label{dd:2}
\end{table}
% latex table generated in R 4.2.2 by xtable 1.8-4 package
% Fri Jan 13 00:27:52 2023
\begin{table}[!h]
\centering
\begin{tabular}{rcccc}
  \hline
  & \multicolumn{2}{c}{SNR = 20} & \multicolumn{2}{c}{SNR = 5}\\ \hline
 & $M_1$ & $M_2$ & $M_1$ & $M_2$ \\ 
  TFBoost(A.1) & 968.39 (64.41) & 596.26 (263.95) & 699.66 (265.11) & 262.33 (196.84) \\ 
  TFBoost(A.2) & 863.20 (168.48) & 351.35 (268.34) & 481.17 (274.79) & 198.20 (172.81) \\ 
  TFBoost(A.3) & 812.06 (216.51) & 318.31 (253.23) & 486.76 (292.74) & 182.57 (154.14) \\ 
  TFBoost(B) & 608.15 (269.46) & 255.58 (226.69) & 318.60 (261.46) & 154.19 (121.87) \\ 
   \hline
\end{tabular}
\caption{Summary statistics of the early stoping times $T_{\text{stop}}$  selected by TFBoost methods for data generated from $r_2$; displayed in the form of mean (sd).} 
\label{e:2}
\end{table}
% latex table generated in R 4.2.2 by xtable 1.8-4 package
% Fri Jan 13 00:27:53 2023
\begin{table}[!h]
\centering
\begin{tabular}{rcccc}
  \hline
  & \multicolumn{2}{c}{SNR = 20} & \multicolumn{2}{c}{SNR = 5}\\ \hline
 & $M_1$ & $M_2$ & $M_1$ & $M_2$ \\ 
  TFBoost(A.1) & 0.220 (0.014) & 0.082 (0.013) & 0.694 (0.041) & 0.160 (0.014) \\ 
  TFBoost(A.2) & 0.216 (0.014) & 0.080 (0.012) & 0.693 (0.037) & \textbf{0.159} (0.014) \\ 
  TFBoost(A.3) & \textbf{0.214} (0.014) & \textbf{0.079} (0.012) & 0.688 (0.035) & 0.161 (0.014) \\ 
  TFBoost(B) & \textbf{0.195} (0.011) & \textbf{0.061} (0.009) & \textbf{0.670} (0.035) & \textbf{0.143} (0.011) \\ 
  FLM1 & 1.358 (0.061) & 0.689 (0.038) & 1.788 (0.086) & 0.759 (0.039) \\ 
  FLM2 & 1.352 (0.065) & 0.689 (0.038) & 1.781 (0.089) & 0.759 (0.039) \\ 
  FAM & 0.947 (0.117) & 0.501 (0.042) & 1.383 (0.129) & 0.572 (0.044) \\ 
  FPPR & 0.237 (0.028) & 0.114 (0.033) & 0.718 (0.056) & 0.201 (0.037) \\ 
  FAME & 0.231 (0.023) & 0.092 (0.021) & 0.720 (0.048) & 0.188 (0.028) \\ 
  FGAM & 0.219 (0.013) & 0.438 (0.029) & \textbf{0.683} (0.036) & 0.512 (0.031) \\ 
  FRF & 0.370 (0.041) & 0.279 (0.037) & 0.783 (0.053) & 0.348 (0.037) \\ 
  RFGroove & 0.459 (0.033) & 0.314 (0.032) & 0.913 (0.051) & 0.386 (0.033) \\ 
   \hline
\end{tabular}
\caption{Summary statistics of test errors for data generated from $r_3$; displayed in the form of mean (sd).} 
\label{er:3}
\end{table}
% latex table generated in R 4.2.2 by xtable 1.8-4 package
% Fri Jan 13 00:27:54 2023
\begin{table}[!h]
\centering
\begin{tabular}{rcccc}
  \hline
  & \multicolumn{2}{c}{SNR = 20} & \multicolumn{2}{c}{SNR = 5}\\ \hline
 & $M_1$ & $M_2$ & $M_1$ & $M_2$ \\ 
  TFBoost(A.1) & 1.16 (0.39) & 1.08 (0.27) & 1.39 (0.62) & 1.06 (0.24) \\ 
  TFBoost(A.2) & 1.81 (0.87) & 1.43 (0.57) & 1.88 (0.95) & 1.47 (0.73) \\ 
  TFBoost(A.3) & 1.80 (0.90) & 1.76 (0.88) & 1.81 (0.90) & 1.71 (0.86) \\ 
  TFBoost(B) & 2.97 (0.88) & 3.12 (0.73) & 2.22 (0.96) & 2.97 (0.88) \\ 
   \hline
\end{tabular}
\caption{Summary statistics of the tree depths selected by TFBoost  for data generated from $r_3$; displayed in the form of mean (sd).} 
\label{dd:3}
\end{table}
% latex table generated in R 4.2.2 by xtable 1.8-4 package
% Fri Jan 13 00:27:55 2023
\begin{table}[!h]
\centering
\begin{tabular}{rcccc}
  \hline
  & \multicolumn{2}{c}{SNR = 20} & \multicolumn{2}{c}{SNR = 5}\\ \hline
 & $M_1$ & $M_2$ & $M_1$ & $M_2$ \\ 
  TFBoost(A.1) & 950.58 (116.79) & 977.61 (58.60) & 582.67 (245.51) & 889.96 (136.17) \\ 
  TFBoost(A.2) & 818.94 (237.53) & 926.64 (134.11) & 459.91 (274.04) & 765.63 (252.55) \\ 
  TFBoost(A.3) & 780.60 (257.62) & 898.08 (158.45) & 441.94 (260.47) & 712.90 (264.68) \\ 
  TFBoost(B) & 304.22 (238.92) & 435.92 (257.85) & 232.23 (181.15) & 231.74 (200.35) \\ 
   \hline
\end{tabular}
\caption{Summary statistics of the early stoping times $T_{\text{stop}}$  selected by TFBoost methods for data generated from $r_3$; displayed in the form of mean (sd).}
\label{e:3} 
\end{table}
% latex table generated in R 4.2.2 by xtable 1.8-4 package
% Fri Jan 13 00:27:57 2023
\begin{table}[!h]
\centering
\begin{tabular}{rcccc}
  \hline
  & \multicolumn{2}{c}{SNR = 20} & \multicolumn{2}{c}{SNR = 5}\\ \hline
 & $M_1$ & $M_2$ & $M_1$ & $M_2$ \\ 
  TFBoost(A.1) & 0.252 (0.020) & 0.271 (0.027) & 0.570 (0.038) & 0.583 (0.037) \\ 
  TFBoost(A.2) & 0.237 (0.019) & 0.262 (0.025) & 0.558 (0.033) & 0.575 (0.035) \\ 
  TFBoost(A.3) & 0.226 (0.019) & \textbf{0.260} (0.026) & 0.550 (0.033) & \textbf{0.574} (0.036) \\ 
  TFBoost(B) & \textbf{0.196} (0.016) & \textbf{0.248} (0.026) & \textbf{0.524} (0.030) & \textbf{0.563} (0.039) \\ 
  FLM1 & 1.438 (0.071) & 0.586 (0.036) & 1.712 (0.080) & 0.857 (0.047) \\ 
  FLM2 & 1.454 (0.071) & 0.586 (0.036) & 1.725 (0.079) & 0.855 (0.045) \\ 
  FAM & 0.898 (0.145) & 0.525 (0.042) & 1.174 (0.148) & 0.795 (0.051) \\ 
  FPPR & \textbf{0.209} (0.024) & 0.261 (0.046) & \textbf{0.540} (0.048) & 0.588 (0.057) \\ 
  FAME & 0.308 (0.044) & 0.339 (0.044) & 0.628 (0.054) & 0.657 (0.059) \\ 
  FGAM & 0.300 (0.022) & 0.290 (0.032) & 0.618 (0.037) & 0.587 (0.042) \\ 
  FRF & 0.368 (0.027) & 0.402 (0.031) & 0.635 (0.037) & 0.662 (0.038) \\ 
  RFGroove & 0.487 (0.037) & 0.481 (0.039) & 0.777 (0.049) & 0.758 (0.049) \\ 
   \hline
\end{tabular}
\caption{Summary statistics of test errors for data generated from $r_4$; displayed in the form of mean (sd).} 
\label{er:4}
\end{table}
% latex table generated in R 4.2.2 by xtable 1.8-4 package
% Fri Jan 13 00:27:58 2023
\begin{table}[!h]
\centering
\begin{tabular}{rcccc}
  \hline
  & \multicolumn{2}{c}{SNR = 20} & \multicolumn{2}{c}{SNR = 5}\\ \hline
 & $M_1$ & $M_2$ & $M_1$ & $M_2$ \\ 
  TFBoost(A.1) & 2.19 (0.79) & 1.98 (0.51) & 2.10 (0.83) & 1.74 (0.66) \\ 
  TFBoost(A.2) & 2.58 (0.71) & 2.40 (0.62) & 2.22 (0.94) & 2.21 (0.91) \\ 
  TFBoost(A.3) & 2.50 (0.69) & 2.47 (0.67) & 2.15 (0.91) & 2.12 (0.89) \\ 
  TFBoost(B) & 3.06 (0.79) & 2.45 (0.69) & 2.85 (0.95) & 2.30 (0.86) \\ 
   \hline
\end{tabular}
\caption{Summary statistics of the tree depths selected by TFBoost  for data generated from $r_4$; displayed in the form of mean (sd).} 
\label{dd:4}
\end{table}

% latex table generated in R 4.2.2 by xtable 1.8-4 package
% Fri Jan 13 00:27:59 2023
\begin{table}[!h]
\centering
\begin{tabular}{rcccc}
  \hline
  & \multicolumn{2}{c}{SNR = 20} & \multicolumn{2}{c}{SNR = 5}\\ \hline
 & $M_1$ & $M_2$ & $M_1$ & $M_2$ \\ 
  TFBoost(A.1) & 962.39 (55.28) & 902.47 (134.21) & 781.78 (243.59) & 772.73 (233.61) \\ 
  TFBoost(A.2) & 920.57 (131.91) & 844.47 (180.19) & 682.31 (299.37) & 603.05 (317.61) \\ 
  TFBoost(A.3) & 870.27 (181.13) & 810.42 (233.13) & 644.36 (304.11) & 647.26 (326.90) \\ 
  TFBoost(B) & 527.76 (263.12) & 653.43 (261.88) & 295.81 (256.69) & 417.91 (298.59) \\ 
   \hline
\end{tabular}
\caption{Summary statistics of the early stoping times $T_{\text{stop}}$  selected by TFBoost methods for data generated from $r_4$; displayed in the form of mean (sd).} 
\label{e:4}
\end{table}

\begin{figure}[!h]
\centering
\includegraphics[scale = 0.5, page = 1]{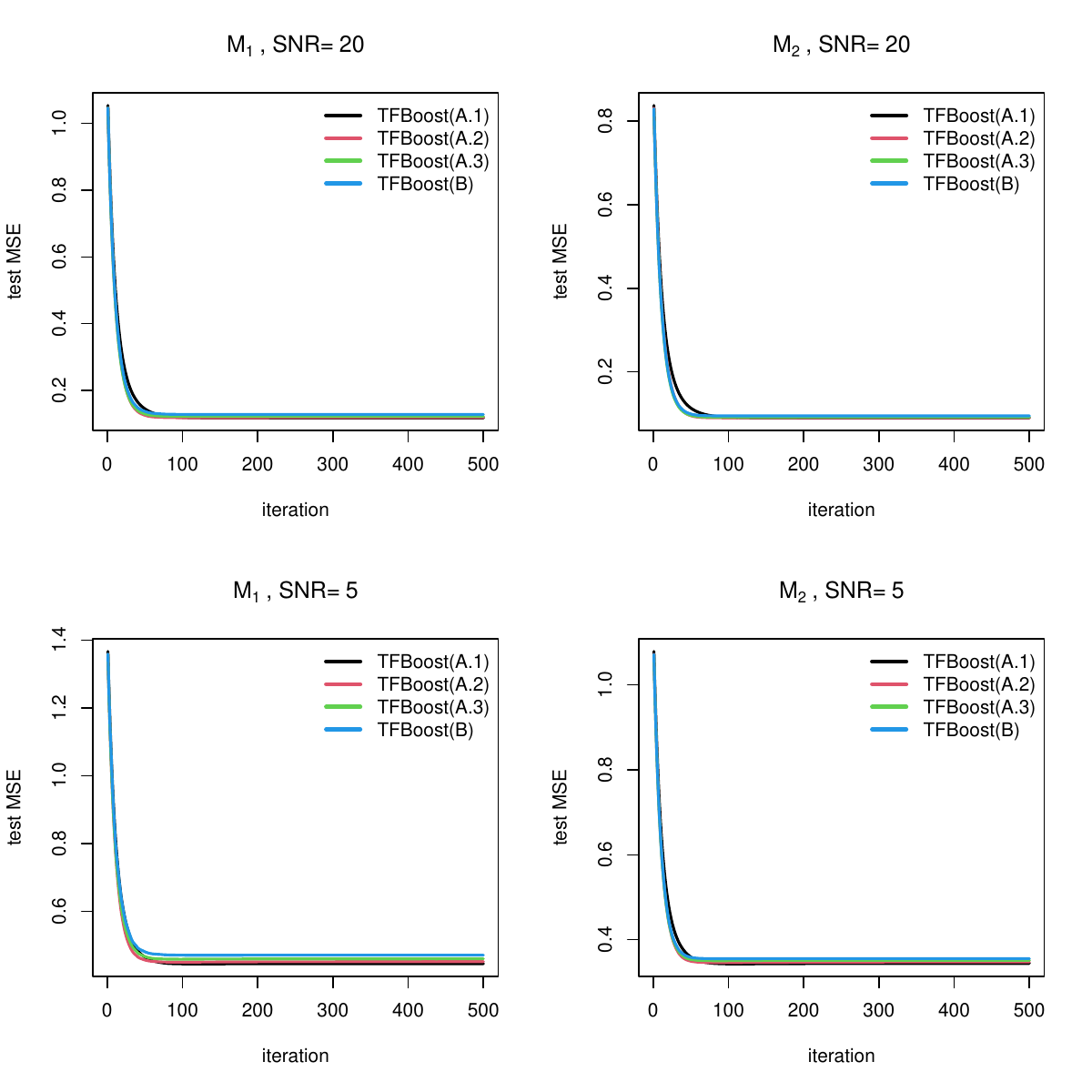}	
\caption{Test MSEs averaged from 100 runs of the experiment for TFBoost in $r_1$ settings}
\label{tr:1}
\end{figure}

\begin{figure}[!h]
\centering
\includegraphics[scale = 0.5, page = 2]{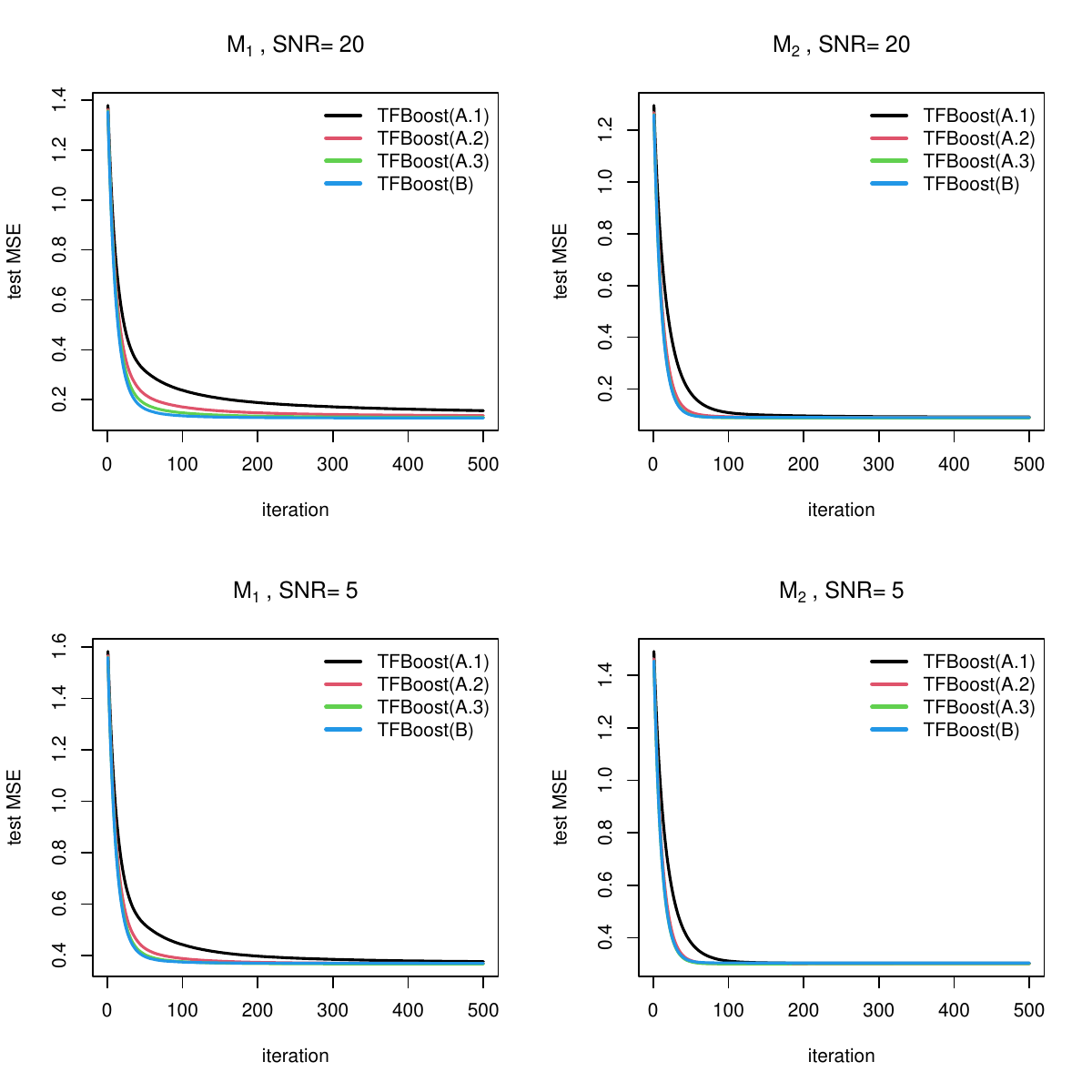}	
\caption{Test MSEs averaged from 100 runs of the experiment for TFBoost in $r_2$ settings}
\label{tr:2}
\end{figure}

\begin{figure}[!h]
\centering
\includegraphics[scale = 0.5, page = 3]{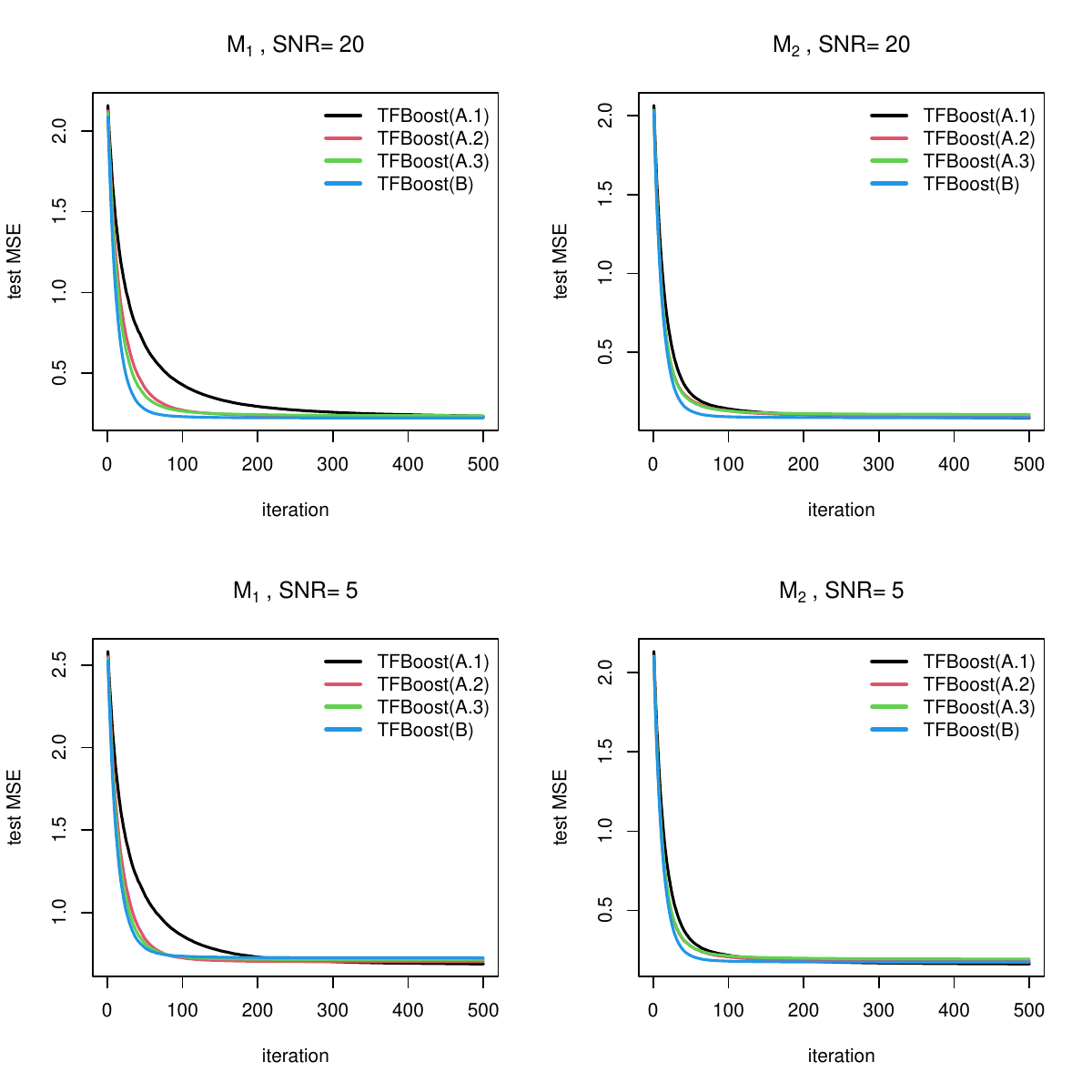}	
\caption{Test MSEs averaged from 100 runs of the experiment for TFBoost in $r_3$ settings}
\label{tr:3}
\end{figure}

\begin{figure}[!h]
\centering
\includegraphics[scale = 0.5, page = 4]{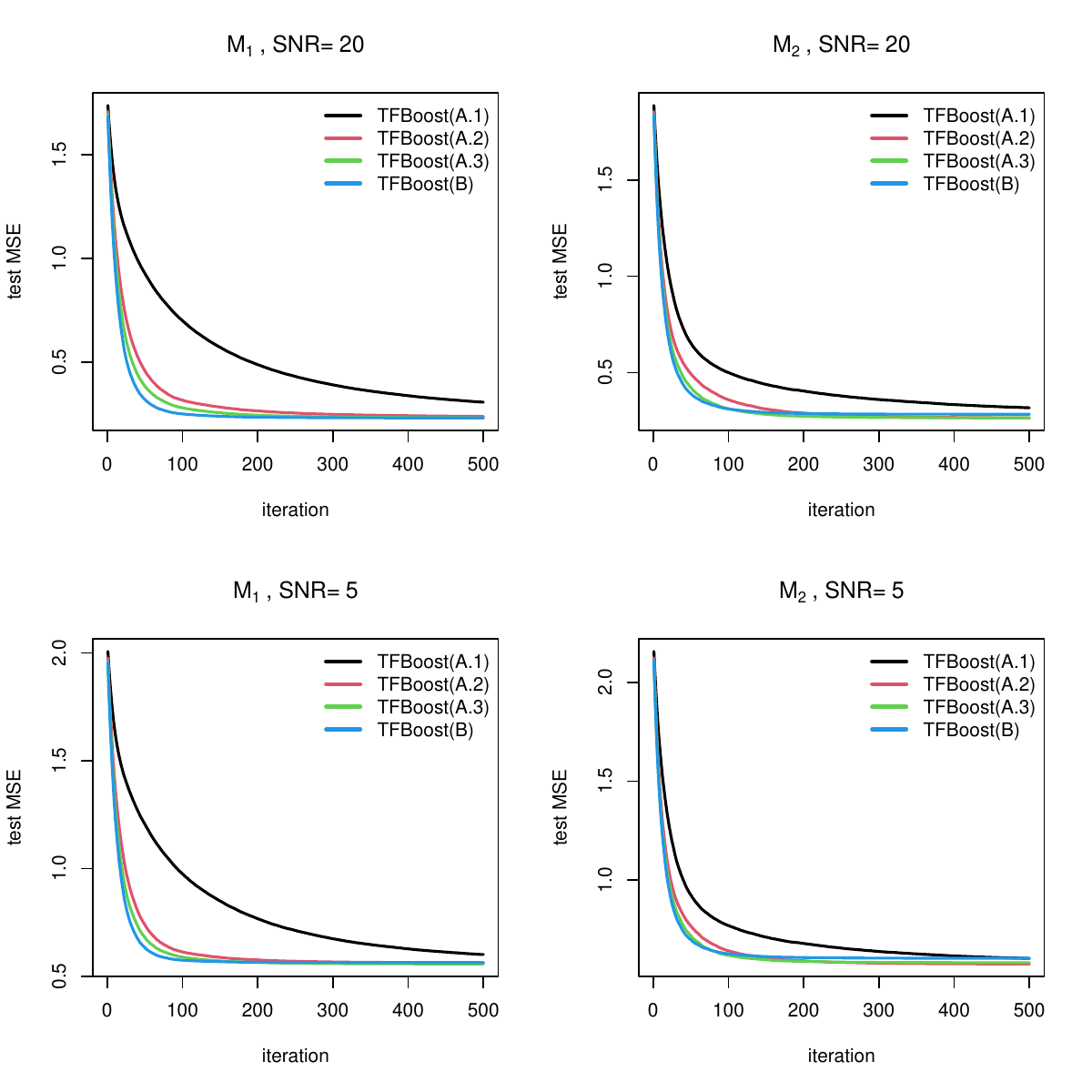}	
\caption{Test MSEs averaged from 100 runs of the experiment for TFBoost in $r_4$ settings}
\label{tr:4}
\end{figure}

\Cref{tr:1,tr:2,tr:3,tr:4} show the  test MSEs averaged over 100 runs of the experiment versus the number of iterations for \texttt{TFBoost.}   For the convenience of taking the averages, for each run, we let the test errors for the iterations past the early stopping time to keep the same test error as the one obtained at the early stopping time.  It can be observed from figures that the test errors usually drop quickly within 100 iterations.

\section*{Appendix C}
We consider another regression function that is linear:
$$r_5(X) =  \int_{\mathcal{I}} \left (\text{sin} \left(\frac{3}{2} \pi t \right) +  \text{sin} \left(\frac{1}{2} \pi t \right)\right)X(t)dt.$$
The other specifications of the model remain the same as described in Section~3. \Cref{er:5} include the summary statistics of test MSEs from 100 independent runs of the simulation, with bold font indicating the lowest two average test errors in each setting.  \Cref{d:5} and \Cref{e:5} include the summary statistics of the tree depths and early stopping times selected by TFBoost methods.

% latex table generated in R 4.2.2 by xtable 1.8-4 package
% Fri Jan 13 00:28:00 2023
\begin{table}[!h]
\centering
\begin{tabular}{rcccc}
  \hline
  & \multicolumn{2}{c}{SNR = 20} & \multicolumn{2}{c}{SNR = 5}\\ \hline
 & $M_1$ & $M_2$ & $M_1$ & $M_2$ \\ 
  TFBoost(A.1) & 0.105 (0.005) & 0.035 (0.002) & 0.415 (0.019) & 0.132 (0.006) \\ 
  TFBoost(A.2) & 0.107 (0.005) & 0.035 (0.002) & 0.419 (0.019) & 0.133 (0.006) \\ 
  TFBoost(A.3) & 0.108 (0.005) & 0.035 (0.002) & 0.418 (0.019) & 0.134 (0.006) \\ 
  TFBoost(B) & 0.109 (0.005) & 0.037 (0.002) & 0.420 (0.020) & 0.136 (0.006) \\ 
  FLM1 & \textbf{0.100} (0.005) & \textbf{0.031} (0.001) & 0.398 (0.018) & \textbf{0.124} (0.006) \\ 
  FLM2 & \textbf{0.099} (0.004) & \textbf{0.031} (0.001) & \textbf{0.395} (0.017) & \textbf{0.123} (0.005) \\ 
  FAM & 0.191 (0.044) & 0.033 (0.003) & 0.490 (0.053) & 0.126 (0.006) \\ 
  FPPR & 0.102 (0.005) & 0.033 (0.002) & 0.407 (0.022) & 0.128 (0.006) \\ 
  FAME & 0.102 (0.005) & \textbf{0.032} (0.002) & 0.408 (0.021) & 0.128 (0.007) \\ 
  FGAM & \textbf{0.099} (0.004) & \textbf{0.031} (0.001) & \textbf{0.396} (0.017) & \textbf{0.123} (0.005) \\ 
  FRF & 0.150 (0.010) & 0.050 (0.004) & 0.442 (0.021) & 0.140 (0.007) \\ 
  RFGroove & 0.150 (0.008) & 0.100 (0.009) & 0.469 (0.021) & 0.194 (0.012) \\ 
   \hline
\end{tabular}
\caption{Summary statistics of test errors for data generated from $r_5$; displayed in the form of mean (sd).} 
\label{er:5}
\end{table}
% latex table generated in R 4.2.2 by xtable 1.8-4 package
% Fri Jan 13 00:28:01 2023
\begin{table}[!h]
\centering
\begin{tabular}{rcccc}
  \hline
  & \multicolumn{2}{c}{SNR = 20} & \multicolumn{2}{c}{SNR = 5}\\ \hline
 & $M_1$ & $M_2$ & $M_1$ & $M_2$ \\ 
  TFBoost(A.1) & 3.03 (0.88) & 3.33 (0.79) & 2.57 (1.07) & 3.05 (0.89) \\ 
  TFBoost(A.2) & 2.63 (0.86) & 3.22 (0.88) & 2.23 (0.97) & 2.61 (1.00) \\ 
  TFBoost(A.3) & 2.34 (0.86) & 3.20 (0.72) & 1.94 (0.81) & 2.60 (0.96) \\ 
  TFBoost(B) & 1.59 (0.71) & 2.25 (1.02) & 1.43 (0.59) & 1.79 (0.78) \\ 
   \hline
\end{tabular}
\caption{Summary statistics of the tree depths selected by TFBoost  for data generated from $r_5$; displayed in the form of mean (sd).} 
\label{d:5}
\end{table}
% latex table generated in R 4.2.2 by xtable 1.8-4 package
% Fri Jan 13 00:28:02 2023
\begin{table}[!h]
\centering
\begin{tabular}{rcccc}
  \hline
  & \multicolumn{2}{c}{SNR = 20} & \multicolumn{2}{c}{SNR = 5}\\ \hline
 & $M_1$ & $M_2$ & $M_1$ & $M_2$ \\ 
  TFBoost(A.1) & 160.91 (98.69) & 186.69 (168.13) & 135.25 (71.70) & 167.59 (104.61) \\ 
  TFBoost(A.2) & 136.71 (88.51) & 154.42 (115.92) & 120.17 (64.89) & 149.28 (97.01) \\ 
  TFBoost(A.3) & 144.23 (87.28) & 143.58 (80.80) & 120.14 (65.11) & 145.25 (111.65) \\ 
  TFBoost(B) & 144.77 (62.02) & 177.38 (125.64) & 115.78 (54.95) & 152.31 (108.28) \\ 
   \hline
\end{tabular}
\caption{Summary statistics of the early stoping times $T_{\text{stop}}$  selected by TFBoost methods for data generated from $r_5$; displayed in the form of mean (sd).}
\label{e:5} 
\end{table}

\section*{Appendix D}
The summary statistics of test MSEs from 100 random partitions of the German electricity data in \Cref{sec:real} are provided  in \Cref{ap:sum}.

\begin{table}[!h]
\centering
\begin{tabular}{rll}
  \hline
 & Part (1) &  Part (2)  \\  
 \hline
TFBoost(A.1) & 5.333 (0.649) & 3.933 (0.527) \\
  TFBoost(A.2) & 5.485 (0.642) & 3.876 (0.492) \\   TFBoost(A.3) & 5.437 (0.602) & 3.924 (0.495) \\  
  TFBoost(B) &  5.004 (0.561) & 4.436 (0.548) \\   
  FLM1 & 10.656 (4.814) & 10.692 (5.040) \\ 
  FLM2 & 12.345 (7.511) & 11.507 (6.017) \\ 
    FAM & 13.069 (16.914) & 9.4057 (17.358) \\ 
    FPPR & 6.224 (0.715) & 5.710 (0.704) \\   
    FAME &  12.646 (13.254) & 9.430 (7.899) \\
  FGAM &  7.409 (2.182) & 5.797 (1.133) \\ 
  FRF & 5.430 (0.681) & 4.519 (0.548) \\ 
  RFgroove & 5.535 (0.636) & 4.589 (0.517) \\ 
   \hline
\end{tabular}
\caption{Summary statistics of test MSEs displayed in the form of mean (sd). The unit of MSEs is $10^{-6}$.}
\label{ap:sum}
\end{table}




\end{document}

\bibliographystyle{apalike}

\def\spacingset#1{\renewcommand{\baselinestretch}%
{#1}\small\normalsize} \spacingset{1}

%%%%%%%%%%%%%%%%%%%%%%%%%%%%%%%%%%%%%%%%%%%%%%%%%%%%%%%%%%%%%%%%%%%%%%%%%%%%%%

\if0\blind
{
  \title{\bf Tree-based boosting with functional data}
  \author{Author 1\thanks{
    The authors gratefully acknowledge \textit{please remember to list all relevant funding sources in the unblinded version}}\hspace{.2cm}\\
    Department of YYY, University of XXX\\
    and \\
    Author 2 \\
    Department of ZZZ, University of WWW}
  \maketitle
} \fi

\if1\blind
{
  \bigskip
  \bigskip
  \bigskip
  \begin{center}
    {\LARGE\bf Functional tree boosting}
\end{center}
  \medskip
} \fi

\input{appendix/val_nknots_3.tex}

\typeout{}
\bibliography{mybibfile}